\documentclass[preprint,12pt]{elsarticle}

\usepackage{amssymb}
\usepackage{amsmath}

\usepackage[hyphens]{url}

\usepackage{hyperref}
\usepackage{cleveref}

\usepackage{tabularray}
\UseTblrLibrary{tikz}

\ExplSyntaxOn
\cs_generate_variant:Nn \clist_map_inline:nn {e}
\cs_new_protected:Npn \mytreepass #1
{
\clist_map_inline:en {\ExpTblrChildClass {#1}}
{
 \exp_args:Noo \mytreepassfill {\use_i:nn ##1} {\use_ii:nn ##1}
}
}
\cs_new_protected:Npn \mytreesub #1
{
\clist_map_inline:en {\ExpTblrChildClass {#1}}
{
 \exp_args:Noo \mytreesubfill {\use_i:nn ##1} {\use_ii:nn ##1}
}
}
\cs_new_protected:Npn \mytreesubend #1
{
\clist_map_inline:en {\ExpTblrChildClass {#1}}
{
 \exp_args:Noo \mytreesubendfill {\use_i:nn ##1} {\use_ii:nn ##1}
}
}
\ExplSyntaxOff

\newcommand\mytreepassfill[2]{
\draw (#1-#2.north) --  (#1-#2.south);
}
\newcommand\mytreesubfill[2]{
\draw (#1-#2.north) --  (#1-#2.south) (#1-#2.mid) -- (#1-#2.mid east);
\draw let \n1={int(#1-1)} in let \p1=(\n1-#2.mid), \p2=(#1-#2) in ([yshift=-5]\x2, \y1) -- (#1-#2.north);  
}
\newcommand\mytreesubendfill[2]{
\draw (#1-#2.north) --  (#1-#2.mid) -- (#1-#2.mid east);
\draw let \n1={int(#1-1)} in let \p1=(\n1-#2.mid), \p2=(#1-#2) in ([yshift=-5]\x2, \y1) -- (#1-#2.north); 
}

\newcommand{\theme}{\SetChild{class=theme,classh=theme}}
\newcommand{\code}{\SetChild{class=code}}
\newcommand{\pass}{\SetChild{class=pass}}
\newcommand{\sub}{\SetChild{class=sub}}
\newcommand{\subend}{\SetChild{class=subend}}

\UseTblrLibrary{counter}  
\usepackage{rotating}
\usepackage{graphicx}
\usepackage{adjustbox}
\usepackage{array}

\usepackage{tabularx}

\usepackage{longtable} 
\usepackage{wasysym}
\usepackage{xcolor,colortbl}
\definecolor{grey}{rgb}{0.727,0.727,0.727}

\usepackage{caption}  
\usepackage{pdflscape} 

\newcolumntype{R}[2]{%
    >{\adjustbox{angle=#1,lap=\width-(#2)}\bgroup}%
    l%
    <{\egroup}%
}
\newcommand*\rot{\multicolumn{1}{R{45}{1em}}}

\newcounter{finding}
\newcommand{\findingsnumber}{\stepcounter{finding}\thefinding}

\usepackage{mdframed}  

\newcommand{\zerowidthfootnote}[1]{\makebox[0pt][l]{\footnotemark}\footnotetext{#1}}

\journal{Nuclear Physics B}

\begin{document}

\begin{frontmatter}



\title{Comparing and Conceptualizing Data Protection Requirements Worldwide for Privacy Regulatory Compliance} 

 \affiliation[label1]{organization={ULIDE, University of Luxembourg},
             country={Luxembourg}}
 \affiliation[label2]{organization={Human-Media Interaction Research Group, University of Twente},
            country={The Netherlands}}

 \affiliation[label3]{organization={FINATRAX, SnT, University of Luxembourg},
             country={Luxembourg}}
             
\affiliation[label4]{organization={IRiSC, SnT, University of Luxembourg},
             country={Luxembourg}}
             
\affiliation[label5]{organization={Escuela de Ingeniería Informática, Universidad de Valparaíso},
             country={Chile}}
\affiliation[label6]{organization={Polytechnic University of Valencia},
             country={Spain}}

\author[label1]{Claudia Negri-Ribalta\footnote{Correspoding author, claudia.negriribalta@uni.lu}}

\author[label2]{Lorena Sanchez Chamorro} 
\author[label1]{Ioana Visescu} 
\author[label3]{Muriel Frank}
\author[label4]{Anastasia Sergeeva} 
\author[label6]{Alberto García} 
\author[label5]{Rene Noel} 


\begin{abstract}
The growing digitalization of society has intensified the collection, processing, and sharing of personal data, increasingly moving across national borders and regulatory jurisdictions, prompting a proliferation of data protection frameworks worldwide. These transborder personal data flows (TPDF) are essential to today's economy, but organizations managing them must reconcile data protection requirements that differ, sometimes subtly, across jurisdictions. For requirements engineering, this is the central challenge: regulatory data protection requirements (RDPRs) are complex and not directly translatable into software requirements, especially when frameworks impose similar, non-identical, or contradictory obligations. Identifying which requirements are shared and which diverge is therefore critical to managing TPDF, and addressing them late in the software development lifecycle (SDLC) causes costly rework, making early identification essential for compliance and stakeholder communication. This paper identifies and conceptualizes common RDPRs worldwide from the perspective of data protection legal experts, answering: (SQ1) which requirements are common across regulations and how are they conceptualized, and (SQ2) which requirements diverge and how do they differ conceptually. We combine deductive qualitative analysis of interviews with 70 legal experts from G20 economies and other countries and systematic content analysis of these economies' data protection regulations. We identify common requirements, such as consent, and divergent ones, such as the right to be forgotten. Given their impact across the SDLC and enterprise architecture, we translate these findings into a set of Data Protection Officer DPO (DPO) stories, using the user story notation, classified by SDLC phase and enterprise architecture layer, to help organizations manage TPDF compliance.
\end{abstract}




\end{frontmatter}




\section{Introduction}

The increasing digitalization of society has led to increased personal data collection, processing, and sharing. With this phenomenon, the usage of resources from organizations across the world, such as servers and software, has increased, leading to complex international data flows \cite{aaditya2019international}. Consequently, the establishment of robust regulatory frameworks worldwide to protect the privacy of individuals and ensure the ethical handling of personal data has proliferated \cite{OECD_crossborder}. Regulatory instruments, such as the European Union's General Data Protection Regulation (GDPR) \cite{GDPR}, California's Consumer Privacy Act (CCPA) \cite{CCPA}, Brazil's Lei Geral de Proteção de Dados (LGPD) \cite{LGPD}), among other national and regional data protection laws, define a wide array of requirements that organizations must adhere to when managing personal data. 

These international data flows across jurisdictions involve different data types, including personal data. The process of sending and processing personal data across regulations is known as  `Transborder Personal Data Flows (TPDF)'. The Organisation for Economic Co-operation and Development (OECD) \cite{OECD_transborder} defines TPDF as `movements of personal data across national borders' and is fundamental for today's economy \cite{aaditya2019international}. However, the diversity of regulatory approaches and their  context-specific requirements present significant challenges for organizations aiming to achieve compliance across jurisdictions, complicating TPDF \cite{aaditya2019international,OECD_crossborder}. The complexities in these regulations make their integration into software systems both essential and challenging \cite{breaux22}.

TPDF represents an open challenge for the requirements engineering (RE) community as it requires continuous efforts to identify and trace regulatory data protection requirements (RDPRs) \cite{negri2024understanding,breaux22}. Moreover RDPRs are complex and are not often immediately understandable by software engineers on how to translate them into software requirements \cite{breaux22,hadar2018privacy}. This complexity increases in the context of TPDF, where multiple regulatory frameworks may impose the same, similar but not exactly equal, or even conflicting obligations \cite{OECD_crossborder}. Finally, incorporating such requirements late in the software development lifecycle often results in significant rework, which leads to increased costs and delays \cite{pohleng}. It is therefore vital to include requirements from early phases of the software development lifecycle (SDLC). 

Developing a unified understanding of commonalities and particularities of these requirements across regulations is vital for the SDLC and compliance efforts in transborder contexts. This shared understanding of RDPRs can help establish a fluid and effective conversation between experts in the law and computer science, or even among other stakeholders \cite{glinz2015shared}. Although certain levels of implicit understanding of requirements may be needed \cite{glinz2015shared}, a defined conceptualization is needed given the legal nature of RDPRs. 

This paper seeks to contribute to these efforts by identifying and conceptualizing common RDPRs worldwide through the lenses of data protection legal experts. In particular, we aim to answer the following questions:
\begin{itemize}
\item SQ1: Which data protection requirements are common across regulations, and how are they conceptualized?
\item SQ2: Which data protection requirements diverge across regulations, and how are they conceptually different?
\end{itemize}

We aim to answer these questions through a mixed-method approach: we interviewed 70 legal experts from around the would which we analyzed with deductive qualitative analysis, and analyzed through systematic content analysis 21 data protection regulation. The interview participants are experts in data protection or privacy, coming from different legal practices (ranging from public officials to private practitioners) with different years of expertise. The data protection regulations we analyze are from the G20 economies and other selected countries. We have identified common requirements, such as consent, and divergent requirements, such as the right to be forgotten/erasure.

These findings may affect not only several phases of the SDLC, but also the whole enterprise architecture (EA) \cite{opengroup_togaf}, from the organization's strategy to its technology infrastructure, across several applications and data management policies, making the operationalization of our findings challenging. To address this, we use the user story notation \cite{cohn2004userstories} (widely adopted in the industry for requirements) and propose a set of \textit{DPO}\footnote{DPO stands for Data Protection Officer.} \textit{stories}, derived from the RDPR findings, classified according to the SDLC phase and the EA layer in which they should be addressed. This artifact should help organizations identify the common requirements across regulations and, perhaps more importantly, the requirements that differ, thus helping organizations manage TPDFs.

The remainder of this paper is structured as follows: we first discuss and provide a theoretical background section on data protection requirements in \Cref{sec:related-work}, followed by related work in TPDFs. Afterwards in \Cref{sec:research-method}, we discuss our methodology. Subsequently, we present our results and findings in \Cref{sec:analysis}, detailing the requirements, commonalities, challenges found, and legal insights elicited from the analysis. We then present and discuss the DPO stories in \Cref{sec:dpo-stories}. Finally, in \Cref{sec:threats} we discuss the limitations of this work and propose future research directions, concluding in \Cref{sec:conclusion}.

\section{Related work} \label{sec:related-work}

\subsection{Privacy principles and international frameworks of data protection}  

From an international perspective, various instruments have been proposed and developed to align privacy and data protection regulations worldwide. Early global efforts include the \textit{`Records, Computers and the Rights of Citizens: Report of the HEW Advisory Committee on Automated Personal Data Systems'} (HEW) \cite{welfare1973records}, which lead to the Fair Information Practice Principles (FIPs)\footnote{It is unclear who first coined the term FIPs \cite{gellman2025fair}.} in the United States; and the European Council’s resolutions, which lead to Convention 108 in 1981 \cite{iapp2019european,council7322}. 

In brief, the HEW report discusses the importance of privacy and  transparency, accessibility, accountability, and purpose limitation, among others as key data protection principles \cite{welfare1973records}. The Federal Trade Commission (FTC) and the Privacy Act of 1974 in the United States adopted this report recommendations as the FIPs \cite{gellman2025fair}, now an internationally recognized standard. In 2008, the Department of Homeland Security revised and adapted the FIPs to eight principles, introducing the acronym FIPPs to differentiate them from the original FIPs \cite{gellman2025fair,teufel2008fair}.

In parallel, various European countries were enacting privacy and data protection laws. The Council of Europe adopted resolutions 73/22 and 74/29 \cite{council7322}, underscoring the importance of privacy and data protection, while setting basic principles and practices (\cite{report1981convention}). By the late 1970s, most Council members had (unharmonized) data protection regulations, affecting TPDF and the idea of an European internal market \cite{reportdirective}. 

Similarly, since 1969, the OECD has launched several initiatives, on the subject \cite{OECDevolving2011}. The \textit{`Transborder Data Flows and the Protection of Privacy'} symposium in 1977 underscored the economic and social importance of TPDF \cite{OECDevolving2011}. Consequently, an ad hoc expert group was established to address the topic and produce privacy guidelines aligned with the OECD and the Council of Europe \cite{OECDevolving2011,OECD_transborder}.

The ad hoc group noted that there was a dichotomy between national laws and the free flow of information, which underscored transborder \cite{OECDevolving2011}. The group also pointed the importance of these regulations for economic growth \citet{OECDevolving2011}. The outcome was a set of generic, non-binding guidelines aimed at protecting individuals' privacy and not just automated processing, known as the OECD privacy principles \cite{OECD_transborder}. The OECD principles became a cornerstone work since they are technology-neutral principles that allow jurisdictions to adapt them to their own idiosyncrasies \cite{OECDevolving2011}. The guidelines have the following principles: collection limitation principle, data quality principle, purpose specification principle, use limitation principle, security safeguards principle, openness principle, individual participation principle, and accountability principle.

Despite their efforts, the OECD and the Council of Europe diverged in developing guidelines, and the Council of Europe ultimately proposed the legally binding Convention 108 (revised as Convention 108+ to align with the GDPR \cite{iapp2019european}) to harmonize data protection laws within its community. The instrument has principles similar to the OECD Privacy Principles, but the terminology and key details differ \cite{OECDevolving2011}. 

From an Asia-Pacific perspective, the Asia-Pacific Economic Cooperation (APEC) first proposed its privacy principles in 2005 \cite{vasquezfrom2025}. These were operationalized in the Cross Border Privacy Rules (CBPR) system in 2011 and expanded globally in 2022 to include non-APEC members, bringing the total to nine economies. In brief, the principles aim to serve as minimum standards among these economies \cite{vasquezfrom2025}. The principles are: preventing harm; notice; collection limitation; use choice; integrity; security safeguards; access and correction; and accountability \cite{APECprivacy2005}. Compared to those from the OECD, they are not that different \cite{vasquezfrom2025}.

The CBPR differs significantly from the GDPR. Indeed, during the adequacy decision on data protection between Japan and the EU, the EU noted in paragraph 79 Japan's membership of the CBPR \cite{JapanAdequacy}. In the decision, in the same paragraph, the EU has indicated that data on EU residents processed in Japan could not be sent to third countries that do not guarantee the same level of data protection \cite{vasquezfrom2025,JapanAdequacy}. Consequently, it has been questioned whether the current state of the Global CBPR is incompatible with the GDPR \cite{vasquezfrom2025}. 

\subsection{Personal data and transborder personal data flows}  

Over the last decade (2015-2026), the number of data protection regulations enacted has increased significantly. According to the UN Conference on Trade and Development (UNCTAD) 79\% of the tracked countries have data protection laws, adding up to more than 144 countries worldwide \cite{iappDataProtectionNumber}. This trend can be explained by three factors. The literature suggests that several factors have contributed to this trend. Among these, the cost of moving data across countries has decreased, making services globally available \cite{OECD_crossborder,jurcys2022future}. 

In this line, academia and specialized economic organizations have recognized the value to society of transborder data flows, including personal data \cite{OECD_crossborder,novotny1981transborder,BABALOLA2024105940,casalini2021,OECD_Enhancing}. To name some benefits, they bring economic benefits, such as the creation of new opportunities, improve the quality of research, and enhance communication between stakeholders \cite{casalini2021,OECD_crossborder,OECD_Enhancing}. However, TPDFs carry risks and challenges, particularly a from regulatory perspective. At times, jurisdictions impose different requirements that conflict with each other, as demonstrated by the so-called ``Schrems'' case \cite{casalini2021}. Sometimes, the same requirement can have a different conceptualization and obligations. For example, the satisfaction of the rights to erasure differs can differ between jurisdictions, as it is the case for the right to be forgotten between Chile and the EU \cite{ortiz2025derecho,ortiz2021repensando}.

This difference in satisfying regulatory requirements demands attention from organizations when they move between jurisdictions, to avoid fines and ensure compliance. Indeed, organizations and stakeholders have expressed their concerns regarding the incompatibility and uncertainty of legal regimes, leading to time-consuming tasks and higher usage of resources \cite{OECD_reportGuidelines,OECD_crossborder,casalini2021}. Therefore, identifying which requirements differ and which are similar can reduce these concerns, which also provides a foundation of RDPRs that can be re-used across jurisdictions.

\subsection{Data protection and privacy requirements} \label{subsec:privacy-dataprotection-requirements}

Translating RDPRs into software requirements presents a significant challenge \cite{breaux22,hadar2018privacy, bednar2019,senarath2018}. Legal texts are written using domain-specific language which makes their translation complex and demands knowledge of other legal instruments \cite{breaux2007systematic}. In addition, software developers frequently perceive privacy requirements as burdensome to implement \cite{bednar2019}, and they deliberately choose not to prioritize them over functional system requirements \cite{senarath2018}. They also perceive them as simply security requirements \cite{hadar2018privacy}, and consider them as `vague' \cite{solove2006taxonomy,bednar2019}, difficult to satisfy \cite{hadar2018privacy,bednar2019,senarath2018}, and less effective in guaranteeing privacy protection than technical (security) measures \cite{sheth2014}.
Given these multiple challenges around understanding and implementing RDPRs, multiple RE artifacts\footnote{We adopt the definition of \citet{wieringa2014design} over artifacts, which are `methods, techniques, notations, and algorithms used in software and information systems [...] [which are] intended to help stakeholders' \cite{wieringa2014design}.} have been proposed \cite{elger2024engenharia,negri2024understanding,gharib2016ontologies} to support their implementation within organizations.
For instance, \citet{agostinelli2019achieving} used BPMNs to propose different patterns for analyzing consent, data breaches, and subjects' rights under the GDPR. Similarly, Unified Modeling Language (UML) has been explored as a possible means of addressing GDPR requirements \cite{torre2021modeling}. From a goal-oriented requirements engineering perspective, the Socio-Technical Security
Modeling Language (STS-ml) framework has been adapted to satisfy GDPR requirements by \citet{robol2017toward} and \citet{negri2023empirical}. \citet{ayal-rivera2018the} proposed the `Guide me' framework, which consists of a six-step systematic approach to eliciting and linking GDPR requirements to privacy obligations. Finally, \citet{esteves2024analysis} carry out a survey of ontologies and vocabularies for the GDPR, concluding that there is no privacy policy language that can model all the GDPR concepts defined by the authors, nor are they necessarily updated to jurisprudence \cite{esteves2024analysis}. Yet most of the work done does not necessarily include legal experts or focuses in just one regulation, such as the GDPR.  

Prior work has explored how to support compliance with data protection requirements in other regions of the world, such as the Global North and Brazil. For example, \citet{breaux2006towards} present a method called semantic parametrizations to extract rules and obligations from the HIPAA regulation, which is further extended and explained in \citet{breaux2008analyzing}. From a Brazilian perspective, \citet{elger2024engenharia} identifies 65 relevant studies for compliance with the LGPD from an RE perspective, concluding that the solutions and artifacts proposed for the GDPR cannot simply be adapted for the LGPD \cite{elger2024engenharia}.This conclusion by \citet{elger2024engenharia} highlights the importance of having artifacts that are jurisdiction agnostic. It also stresses how artifacts for one regulation may not work in another.

In line with \citet{hadar2018privacy,bednar2019,senarath2018}, the findings of \citet{canedo2022guidelines} suggest that security principles are the most well-known aspect from the LGPD (90\% of respondents knew them), followed by transparency (58,5\%), and that practitioners lack the appropriate tools to comply with the regulation \cite{canedo2022guidelines}. Finally, \citet{canedo2021agile} found that 90\% of the agile teams would use user stories for RDPRs, suggesting that this type of artifact is a promising way to operationalize RDPRs in practice, and motivates this works' development of DPO stories.

\subsection{Requirements engineering and user stories} \label{subsec:user-stories} 
A user story is a short, plain-language sentence that captures a single piece of software
functionality from the point of view of the person who needs it. Unlike a formal specification, it is written using everyday words rather than technical jargon, so that both the people who request a feature (for example, a customer or manager) and the people who build it (the software development team) can understand and discuss it easily.  User stories typically follow a three-part template : \textit{``As a \texttt{<role>}, I want a  \texttt{<goal>}, so that \texttt{<reason>}.'' } \cite{cohn2004userstories}.

The \texttt{<role>} identifies
who benefits from the feature, the \texttt{<goal>} states what that person wants the software to do, and the \texttt{<reason>} explains why it matters to them. For instance, a user story for an online store might read: ``As a returning customer, I want to save my delivery address, so that I can check out faster next time.'' Because they are short, concrete and free of technical detail, user stories are commonly used in agile software development to organize a team's list of pending work (the backlog) and to keep the conversation about requirements centered on the people the software is meant to serve, rather than on the software's internal workings.
 
User stories are a widely adopted requirements notation in agile software development teams \cite{lucassen2016improving,kassab2014empirical}. Even though user stories were originally developed to represent the perspective of the software system's users, they
are often also used to express the concerns of other stakeholders or roles in the software product or process~\cite{wautelet2014unifying}. For instance,
technical stakeholders may use them to set engineering constraints, while sponsors may use them to impose process or budget restrictions.

To deliver high-quality user stories, multiple frameworks have been proposed for their development. One that stands out within the RE domain is the AQUSA framework, with a well-developed tool and set of publications \cite{lucassen2016improving,dalpiaz2025automated,8491182,lucassen2015forging}. The AQUSA framework proposes three main criteria: syntactic, semantic, and pragmatic, where each is subdivided into different criteria \cite{lucassen2015forging}. Overall, it proposes 14 criteria for assessing the quality of user stories. Higher-quality user stories enable better understanding of requirements, and hence SDLC. 
 
Several research efforts have adopted user stories to elicit requirements, including legal and regulatory texts, extending the template beyond the end-user role, including privacy and data protection. \citet{alsaadi2019minimizing} rephrase legal requirements from the EU Medical Device Regulation and HIPAA into user stories to remove ambiguity in legal language, building on earlier work by \citet{islam2010towards} and \citet{jorshari2012extracting} on extracting security and privacy requirements from laws and regulations. \citet{seeba2025evaluating} apply a similar pipeline to the EU NIS2 Directive, identifying six personas, none of them the classical end user, and deriving user stories from their goals and dependencies using a goal-oriented model.
 
For RDPRs in the EU, \citet{bartolini2019gdpr} propose `Data Protection backlogs' of GDPR-based user stories mapped to access control policies. At the platform level, the EU DEFeND project elicits and consolidates GDPR compliance requirements from stakeholders across multiple  sectors~\cite{tsohou2020privacy}. Both of these research outputs focus solely in the GDPR. Differently, \citet{peixoto2022evaluating} research the LGPD and propose the Privacy Criteria Method as an alternative to user stories for specifying privacy requirements. In a similar fashion, \citet{canedo2021agile} examine agile teams' perception of privacy requirements elicitation under the LGPD. In fact \citet{canedo2021agile} concluded that the usage of user stories for RDPR by agile teams to be extremely common (90\% of them), as already discussed. 
 
However, across this body of work, data-protection-specific roles,
including the Data Protection Officer, are treated mainly as information
sources for requirements elicitation, rather than as first-class actors
represented directly in the user story's role slot. In this paper, we
build on this line of work by proposing user stories that explicitly
represent data protection and privacy stakeholders, including, but not
limited to, the DPO, as the \texttt{<role>} element of the story itself.

\subsection{Enterprise Architecture and Data Privacy Requirements} \label{subsec:dp-ea}

Data protection and privacy obligations rarely stay within a single system: a requirement such as data minimization or erasure typically involves a policy decision, a change to data models and retention, an application feature, and an infrastructure-level guarantee, touching the organization from governance down to infrastructure. Enterprise architecture (EA) is built to describe exactly this kind of organization-wide concern, decomposing the enterprise into interdependent layers so that requirements can be traced from strategy to infrastructure. The decomposition popularized by TOGAF splits Business, Data, Application, and Technology architecture (``BDAT''): the Business layer captures goals, processes, and policy; the Data layer captures information structure and governance; the Application layer captures software features and workflows; and the Technology layer captures the underlying infrastructure~\cite{opengroup_togaf}.

Existing agile treatments of data protection anchor stories to a single layer: Bartolini et al.'s ``Data Protection backlogs'' render GDPR provisions as access-control requirements at the application layer, dropping the business, data, or technology implications the same requirement carries~\cite{bartolini2019gdpr}. We instead propose a broad set of DP stories spanning the organization and classify each one by the EA layer it belongs to, as other initiatives connect their proposals DP requirements proposals with EA~\cite{burmeister2019privacy, teixeira2021enterprise}. This keeps each story actionable, since each layer maps onto the stakeholders positioned to act on it: compliance and process owners at the business layer, data stewards at the data layer, development teams at the application layer, and infrastructure and security teams at the technology layer.

\section{Research Method} \label{sec:research-method}  

The objective of the study is to identify commonalities across data protection regulations worldwide and to provide conceptualization of these requirements through user stories to support software engineers and to facilitate compliance in TPDF. Thus, this research adopts a deductive approach, starting from previous data on privacy and data protection and existing theory \cite{fife2024deductive,pearse2019illustration}. We aim at discovering patterns, conceptualizations, and refining ongoing theories as part of deductive qualitative approach (DQA) \cite{fife2024deductive,boyatzis1998transforming}. 

\subsection{Participants and recruitment}

We conducted semi-structured interviews with legal experts specializing in data protection. This research received approval by the Ethics Committee of the University of Luxembourg, code ERP 24-109 OBI-PIA. The interviewees were invited using a purposive, non-probabilistic sampling method, focusing on maximum variation in their professional activities as legal experts, including advocacy, government, academia, and the professional world. This strategy was adopted to ensure as much diversity of opinion among the experts \cite{fife2024deductive}. We followed a similar approach as \citet{fereday2006demonstrating} for our sampling. 

While all countries were of interest, the scope was reduced to the G-20 countries\footnote{We excluded Russia from our sample due to the sanctions in place when we received ethical approval.} due to a theoretical reason: they are the leading economies. However, to avoid underrepresentation of specific regions,  
we also include seven selected countries out of those previously invited to the G-20 summits. In these randomly selected countries, at least one country from South America, Asia, and Africa is present. Among addition, we included Luxembourg 
(the country where the project is being carried out). This resulted in a final set of 26 countries of interest, presented in the \Cref{app:list-countries}. 

For countries outside the EU, we sought to recruit at least three experts per country, and where feasible, four or five experts to support thematic saturation \cite{guest2020simple}. For countries within the EU, three experts per country were recruited. This sampling strategy was expected to achieve at least 90\% thematic saturation based on prior qualitative work \cite{guest2020simple}. Overall, observed saturation levels were consistent with previous findings whereby 70\% of codes emerge within the first six interviews and around 92\% within the first twelve interviews \cite{guest2020simple}.

Experts who accepted the invitation to participate were informed about the nature of the study and explained their rights. They were also presented with an information notice, a privacy policy and a consent form via email before the interview. The interviewee signed the documents and the researchers kept a copy.  
The interviewees were offered the opportunity to make their interview transcriptions publicly available, although most consented to keep them confidential. No compensation was offered for participation, which was explained in the invitation email, but all participants will be provided access to the results and materials produced from the research.  

The interviews were offered to be conducted in three possible languages: English, Spanish, or French. The interview questions are available at our public repository\footnote{We provide a Data Availability Statement \Cref{sec:data-availability-statement} for accessing all the publicly available data.}.
Due to the geographical scope of the research, most interviews were conducted online. For best practices, the first 10 minutes (unrecorded and not taken into consideration in this research) were spent building rapport and answering questions. 

The interviews were audio-recorded (when consented) and, with the interviewee's consent, transcribed using AIKO. Transcriptions were anonymized, and the audio file was kept safely processed for the retained period of time defined (three months) and afterwards destroyed, following best practices \cite{strandberg2019ethical}.

\subsection{Data Analysis} \label{subsec:data-analysis}

We analyzed the transcripts following a deductive qualitative analysis (DQA) \cite{fife2024deductive}.  

The DQA approach is a mixed method approach that leverages both deductive and inductive approaches with the purpose of theory testing, refinement or reformulation \cite{fife2024deductive}. 
We provide an overview of the five step method \cite{fife2024deductive}, which we explain in more detail on how we applied afterwards:
\begin{enumerate}
    \item Defining the research questions and the theoretical framework. The research is expected to be positioned within a theoretical framework in order to test their question, concepts, aligned with the deductive approach \cite{boyatzis1998transforming,gilgun2019deductive}. The approach does not require the framework to be formalized \cite{boyatzis1998transforming}, but to be grounded in previous work \cite{gilgun2019deductive} or a conceptual framework \cite{pearse2019illustration}. Our work is based on the GDPR \cite{GDPR}, international comparison of data protection regulations \cite{casalini2021}, and previous work in RE \cite{hadar2018privacy,glinz2015shared,sutcliffe2010collaborative,negri2024understanding}. Our theoretical base is explained in \Cref{sec:related-work} and in further detail in this section.
    \item `Operationalizing Theory: Sensitizing Constructs and Working Hypotheses' \cite{fife2024deductive}.  Our concepts span from the selected framework \cite{gilgun2019deductive} which is the GDPR \cite{GDPR} and can be refined or modified iteratively to better suit the data \cite{fife2024deductive,gilgun2019deductive}. We familiarized ourselves with the code as DQA recommends \cite{fife2024deductive}, to reformulate codes. We explain this procedure in more details when discussing the development of our codebook. 
    \item Sampling. Given the type of research DQA works with (qualitative studies), purpose sampling is carried \cite{fife2024deductive}. This type of sampling is common within qualitative method, due to the type of phenomena and questions asked \cite{guest2006many}. 
    \item `Coding and Analyzing' \cite{fife2024deductive}. The data is coded accordingly, applying the codes derived from the theory as means for analysis. Not only do researchers code and analyze the data, grouping the codes in themes \cite{fife2024deductive}, they also contradict and refute their hypothesis, to reformulate their theory \cite{gilgun2019deductive,fife2024deductive}. 
    \item  Theorizing. The findings, summary of codes, and relationship between them is presented to contribute to the theoretical approach propose \cite{gilgun2019deductive,fife2024deductive}. As part of the DQA tradition, the idea is to challenge and refute original concepts, if the data allows to do so \cite{gilgun2019deductive}.

\end{enumerate}

Applying the aforementioned DQA method, (step 1) we framed our research within the theoretical context of the GDPR \cite{GDPR} in conjunction with previous work from RE \cite{hadar2018privacy,glinz2015shared,sutcliffe2010collaborative,negri2024understanding}. 
To develop the codebook (step 2) we followed the best practices of \citet{boyatzis1998transforming}. They were created using the GDPR \cite{GDPR} data protection requirements \cite{gilgun2019deductive}. While the initial draft was strongly based on the GDPR, it also included further constructs built on identified previous works and grey literature (e.g., the International Association of Privacy Professionals \cite{iapp2019european} work and the OECD \cite{OECD_crossborder}). This initial codebook was discussed with five data protection experts from different continents (Spain, India, Chile, United States and China). Afterwards, the first author initiated an early analysis of the deductive codes, in line with the DQA method \cite{fife2024deductive} in order to immerse into the data and alter the constructs.

For this research, we conceptualize sampling saturation (step 3) as no new more data being discovered, compared to theoretical sampling \cite{hennink2022sample,guest2006many}, following a thematic exhaustion rather than theoretical \cite{guest2006many}. Out final sample is discussed in \Cref{sec:analysis}. 

Through four iterative workshops (step 4), the authors refined the codebook constructs and included new codes using an inductive approach, per allowed as DQA, as part of the sensitizing of our codebook. Once consensus was reached, we split the data among the authors and double-coded the interviews, with each author coding between 10 to 30 interviews and then compared their results, until reaching consensus similarly to \cite{fischer_visescu_devathasan_damian_guzman_2026,larusdottir2024ucd}. All interviews were coded using the specialized software MAXQDA, with an Academic license. We ultimately present our findings by themes that contribute to strengthening previous work and theories in the field (step 5).

\subsection{Regulation notation and legal reviews} 

In order to triangulate the interview findings, we performed a document analysis of the data protection regulations applicable to the countries included in our sample. To identify the relevant regulations we examined the regulatory frameworks referenced by the experts. If the experts did not mention a regulatory framework of reference within their country, we looked into the UNCTAD database \cite{unctadGlobalCyberlaw} and IAPP resources. The purpose of this analysis was not to perform a legal assessment of each regulation, but rather it was intended to support our efforts to identify the existence of specific concepts and their conceptualizations across jurisdictions.

To conduct this analysis, we used systematic content analysis (SCA) for legal analysis \cite{hall2008systematic,brook2022politics,salehijam2018value}. SCA for legal analysis is an adapted form of content analysis used by social scientists \cite{hall2008systematic,salehijam2018value}. It follows a similar pattern: define the research question, collect and code data, analyze, draw conclusions, and report \cite{salehijam2018value,babbie2020practice,hall2008systematic}. 

As such, SCA is particularly suitable for understanding how the law is administered in practice, while also interacting with other areas of research \cite{brook2022politics}. As part of the coding process, the objective is to not only analyze the manifest code, but the latent content, the underlying meaning of what is analyzed \cite{brook2022politics,babbie2020practice,hall2008systematic}. Consequently, SCA in legal research does not aim to replace the interpretation of laws nor achieve normative goals, but rather provides more evidence to a research conclusion \cite{hall2008systematic}.

The coding proceeds similarly to SCA in social science; the coding scheme can be varied, depending on the final objective \cite{hall2008systematic,brook2022politics}. In this article, the codes used come from the same codebook as the one employed for the interviews. As such, we would note where the codes appear in a country's main data protection regulations (and other related laws, as noted by the experts) and keep a record of the articles that mention these codes. This coding scheme allows us to identify trends across regulations and compare the results with the interview data. To further guide our coding of the legal regulations, we supported our analysis with documents from the local data protection authority, academic literature on the country's regulations, and grey literature from data protection professional associations. 

For example, we annotated the following excerpts of the main data protection regulations of Brazil, China, and Indonesia correspondingly as `right to rectification':

\begin{quote}
    
\textit{`Data subjects shall have the right to obtain from the controller, regarding their data processed by said controller, at any time and upon request: correction of incomplete, inaccurate or outdated data'} (Art. 18.III of \cite{Brazil2018LGPD} in English of \cite{LGPD});

\textit{`Where an individual finds that his/her personal information is inaccurate or incomplete, he/she is entitled to request the personal information processor to make corrections or supplements.} (Excerpt of unofficial Chinese-English translation of Art. 46 of \cite{PIPL});

\textit{`A Personal Data Subject shall have the right to complete, update, and/or rectify errors and/or
inaccuracy of Personal Data concerning him/her according to the purpose for which the Personal Data is processed’} (Art. 8, Unofficial English translation of Art. 8 of \cite{PDPLaw})

\end{quote}

Because DQA allows inductive codes, we added codes for elements that had not appeared in the interviews but were in line with requirements identified in the grey literature on the GDPR \cite{GDPR,IAPPComparison}. For example, we created a new code documentation requirement named `Documentation required and registration’ with sub-codes such as privacy policy, privacy notice, data breach log, among others.  

\subsection{User stories for operationalization} \label{subsec:user-stories-methods} 

Based on the findings and conceptualization we arrive based on our data, we develop a set of user stories intended to operationalize these requirements for data protection officers or practitioners. We refer to this artifact as DPO stories because they are designed to support RDPR activities, as our main intended end-user are DPOs. The development of the artifact follows a similar approach to that proposed by \citet{seeba2025evaluating}.

Once the requirements have been elicited and identified, teams develop user stories to implement the requirements \cite{lucassen2016improving}. In this research, the first section corresponds to the requirements elicitation phase, conducted through SCA of the legal text and DQA of the interviews. This elicitation leads to the identification and conceptualization of common RDPRs worldwide, which we present as findings in our discussion. \citet{seeba2025evaluating} follow a similar approach, where they analyze the regulation (NIS2) with a specific framework, to elicit requirements and then transform them into user stories. Based on the data we gathered, we produce DPO stories that should help data protection experts, regardless of jurisdiction, start discussions about RDPRs. 

 The users stories (or DPO stories) follow the classical approach already discussed in \Cref{subsec:user-stories}. For the construction of the DPO stories, we used the AQUSA framework \citet{lucassen2016improving}, as presented in \Cref{subsec:user-stories}. This framework is well-established and recognized in the RE domain. As such, we verify that the DPO stories follow the metrics defined by the framework. For this, we take a two-fold approach: we use the AQUSA-core tool to verify the parameters it allows as provided by \citet{lucassen2016improving}. After multiple iterations with the tool, we have only a couple of warnings across a couple of user stories, which are available in our repository.

As the AQUSA-core tool does not support all the metrics, three authors of this article verified the other metrics in conjunction. As user stories are commonly engineering artifacts used in the SDLC, they do not necessarily follow a scientific method for their development. In contrast, there are metrics and frameworks to measure their quality, such as AQUSA \cite{lucassen2015forging,lucassen2016improving}. However, as part of our contribution, we provide a preliminary validation of these user stories through role-playing, and we aim to provide full validation in future research with real practitioners. 

The lead author, who oversaw the creation of the first round of DPO stories, verified them independently. Afterward, three authors took part in a role-playing poker game\footnote{Poker games are commonly used in agile-based development frameworks that use user stories.} to analyze the DPO stories and modify them accordingly. The lead author served as the domain expert, with the other two serving as the product owner and the architectural engineer, respectively. As stated in our positionality statement (available in \Cref{app:positionality-statement}), the authors of this paper have experience in the domain and have done this exercise in a professional context. During the poker games, the authors paid special attention to the metrics that the AQUSA-core tool did not support. The set of user stories was modified based on the discussions had by the three authors. As part of the discussion, the authors also provided preliminary classifications of the DPO stories, both for the SDLC and the EA, that should help practitioners identify where this requirements gains importance. Future research should focus on providing a more robust classifications of these user stories, applying them in real-life settings.

\section{Data results and analysis} \label{sec:analysis}

We present our results divided by the research questions and subdivided by themes of analysis, providing findings for each theme. The DPO stories are also divided based on the results; namely, those that are common and uncommon, and subdivided by themes. 

\subsection{Semi-structured interviews}
Our final sample was of 74 individuals. We sent out 267 invitations for participation, 74 subject participated, 11 subjects declined, and we had 14 no-shows. 4 subject participated in the interviews, but did not consent (or changed their consent) on their preference on the interview being processed for this research, so their data was not included. The years of expertise in the topic ranged from 2 years to +30, with a work experience of minimum 2 years.

\begin{table} 
\scriptsize
\centering
\SetTblrInner{rowsep=0.25pt} 

\begin{tblr}{
    colspec={cccccc},
    hline{2} = {2-Z}{},
}
    \SetCell[r=2]{c,m} {Num. of \\ Participants}& \SetCell[c=5]{c} Region && \\ \cmidrule[lr]{2-6}
     & Africa & Asia and Oceania & Europe & North America & South America \\ \hline
     Expert &7 & 13 & 15 & 9 & 12 \\
     Intermediate & 3 & 7 & 0 & 0 & 3\\
     Junior & 0 & 1 & 0 & 0 & 0\\
\end{tblr}
\caption{Summary of the participants expertise, based on continent of residence. Junior = 2 - 3 years experience; Intermediate = 4 - 8 years experience; Expert = +8 years experience.} \label{tab:summary-participants}
\vspace{-2em}
\end{table}



\begin{table*}

    \centering
    \scriptsize
    
    \begin{tblrtikzabove}
        \mytreesub{sub}
        \mytreesubend{subend}
        \mytreepass{pass}
    \end{tblrtikzabove}

\SetTblrOuter{expand=\theme\code\pass\sub\subend}
    \makebox[\textwidth]{\begin{tblr}{
        colspec={Q[l,m,1cm]@{\,}Q[l,m,1cm]@{\,}Q[l,m]@{}Q[r, m]Q[l,m]Q[r, m]Q[l,m]Q[r, m]Q[l,m]Q[r, m]Q[l,m]Q[r, m]Q[l,m]Q[r, m]Q[l,m]}, 
        columns = {colsep=2pt},
        cell{1}{1,2,3} = {r=2}{m},  
        column{Z,Z[2]}={gray9,font=\bfseries},  
        hline{2} = {4-Z}{},  
        cell{theme} = {c=3}{l},
        cell{code} = {c=2}{l},
        hspan=even,
        rowsep={0.25pt},
        }
     Theme & Code & Subcode & \SetCell[c=12]{c} {Number of occurrences (percentage of n in column)}  &&&&&&&&&&& \\   
        &&& \SetCell[c=2]{c}{Asia, Oceania \\ n = 21} && \SetCell[c=2]{c}{Africa \\n = 10} && \SetCell[c=2]{c}{N. America\\ n = 9} && \SetCell[c=2]{c}{Europe \\ n = 15} && \SetCell[c=2]{c}{S. America \\n = 15} && \SetCell[c=2]{c}{Total \\n = 70}& \\ 

    \theme  \hline Personal data & &   & 20 & (95.2\%) & 9 & (90.0\%)  & 7 & (77.8\%) & 14 & (93.3\%) & 14 & (93.3\%) & 64 & (91.4\%)      \\

                   \sub & \code Identifiable &  & 15 & (71.4\%) & 4 & (40.0\%)  & 7 & (77.8\%) & 12 & (80.0\%) & 13 & (86.7\%) & 51 & (72.7\%) \\

                   \sub & \code Identified & &  10 & (47.6\%) & 3 & (30.0\%)  & 5 & (55.6\%) & 8 & (53.3\%) & 13 & (86.7\%) & 39 & (55.7\%) \\

                   \pass & \subend & Co-occurrences  & \SetCell[c=2]{c}- && \SetCell[c=2]{c} -  && \SetCell[c=2]{c}- && \SetCell[c=2]{c}- && \SetCell[c=2]{c} - && 38 & (54.3\%) \\

                   \sub & \code Natural persona &   &  7 & (33.3\%) & 3 & (30.0\%)  & 3 & (33.3\%) & 4 & (26.7\%) & 6 & (40.0\%) & 23 & (32.9\%) \\

                   \pass & \subend & Alive  & 4 & (19.1\%) & 0 & (00.0\%)  & 0 & (00.0\%) & 1 & (06.7\%) & 0 & (00.0\%) & 5 & (07.1\%) \\  

                   \sub & \code Legal persona &   & 0 & (00.0\%) & 2 & (20.0\%)  & 0 & (00.0\%) & 0 & (00.0\%) & 0 & (00.0\%) & 2 & (02.9\%) \\  

                  \sub & \code Special categories/sensitive &   &  7 & (33.3\%) & 4 & (40.0\%)  & 4 & (44.4\%) & 2 & (13.3\%) & 5 & (33.3\%) & 22 & (31.4\%) \\

                 \sub & \code Broad/extensive &   &  4 & (19.1\%) & 1 & (10.0\%)  & 1 & (11.1\%) & 3 & (20.0\%) & 1 & (06.7\%) & 10 & (14.3\%) \\

                 \subend & \code The same as the GDPR & & 9 & (42.9\%) & 4 & (40.0\%)  & 1 & (11.1\%) & 3 & (20.0\%) & 3 & (20.0\%) &  20 & (28.6\%) \\
                 \\ 

            \theme {Actors} & &   & 20 & (95.2\%) & 10 & (100\%)  & 6 & (66.7\%) & 13 & (86.7\%) & 15 & (100\%) & 64 & (91.4\%) \\ 
                \sub & \code Data Subject (individual) &  & 7 & (33.3\%) & 7 & (70.0\%) & 1 & (11.1\%) & 7 & (46.7\%)& 8 & (53.3\%) & 30 & (46.9\%) \\
                \sub & \code Controller &  & 15 & (71.4\%) & 9 & (90.0\%) & 4 & (44.4\%) & 8 & (53.3\%)& 12 & (80.0\%) & 48 & (68.6\%) \\
                \sub & \code Processor &  & 15 & (71.4\%) & 7 & (70.0\%) & 4 & (44.4\%) & 6 & (40.0\%) & 11 & (73.3\%) & 43 & (61.4\%) \\
                \sub & \code Data Protection Authority &  & 10 & (47.6\%) & 4 & (40.0\%) & 2 & (22.2\%) & 8 & (53.3\%)& 9 & (60.0\%) & 33 & (47.1\%) \\
                \sub & \code Data Protection Officer\textsuperscript{\dag} &  & 0 & (00.0\%) & 2 & (20.0\%) & 0 & (00.0\%) & 3 & (20.0\%)& 8 & (53.3\%) & 13 & (18.6\%) \\
                \sub & \code Third Parties &  & 1 & (04.8\%) & 1 & (10.0\%) & 0 & (00.0\%) & 3 & (20.0\%)& 3 & (20.0\%) & 8 & (11.4\%) \\
                \sub & \code Joint-controllers\textsuperscript{\dag} &  & 4 & (19.1\%) & 0 & (00.0\%) & 0 & (00.0\%) & 1 & (06.7\%)& 1 & (06.7\%) & 6 & (08.6\%) \\
                \subend & \code The same as the GDPR &  & 3 & (14.3\%) & 3 & (30.0\%) & 0 & (00.0\%) & 1 & (06.7\%) & 0 & (00.0\%) & 7 & (10.0\%) \\
                \\

            \theme {Data Subject Rights} & &   & 18 & (85.7\%) & 9 & (90.0\%)  & 7 & (77.8\%) & 11 & (73.3\%) & 15 & (100\%) & 60 & (85.7\%) \\ 
                \sub & \code Access &  & 12 & (57.1\%) & 3 & (30.0\%) & 5 & (55.5\%) & 5 & (33.3\%)& 10 & (66.7\%) & 35 & (50.0\%) \\
                \sub & \code Rectification &  & 12 & (57.1\%) & 2 & (20.0\%) & 4 & (44.4\%) & 5 & (33.3\%)& 12 & (80.0\%) & 35 & (50.0\%) \\
                \sub & \code Erasure &  & 8 & (38.1\%) & 3 & (30.0\%) & 3 & (30.0\%) & 5 & (55.5\%)& 11 & (73.3\%) & 31 & (44.3\%) \\
                \sub & \code To be Informed &  & 9 & (42.9\%) & 2 & (20.0\%) & 4 & (44.4\%) & 2 & (13.3\%)& 3 & (20.0\%) & 20 & (28.6\%) \\
                \sub & \code Data Portability &  & 5 & (23.8\%) & 3 & (30.0\%) & 2 & (22.2\%) & 4 & (26.7\%)& 3 & (20.0\%) & 17 & (24.3\%) \\

                \sub & \code To Object &  & 3 & (14.3\%) & 2 & (20.0\%) & 1 & (11.1\%) & 3 & (20.0\%)& 5 & (33.3\%) & 14 & (20.0\%) \\

                \sub & \code To Revoke Consent\textsuperscript{\dag}  &  & 3 & (14.3\%) & 2 & (20.0\%) & 1 & (11.1\%) & 2 & (13.3\%)& 5 & (33.3\%) & 13 & (18.6\%) \\

                \sub & \code To Restrict Processing &  & 3 & (14.3\%) & 1 & (10.0\%) & 0 & (00.0\%) & 1 & (06.7\%)& 6 & (40.0\%) & 11 & (15.71\%) \\

                \sub & \code {No to be subject to\\ Automated Decision-Making} &  & 3 & (14.3\%) & 0 & (00.0\%) & 2 & (22.2\%) & 1 & (06.7\%)& 4 & (26.7\%) & 10 & (14.3\%) \\
                \sub & \code `They exist' &  & 2 & (09.5\%) & 3 & (30.0\%) & 2 & (22.2\%) & 3 & (20.0\%)& 7 & (46.7\%) & 17 & (24.3\%) \\

                \subend & \code The same as the GDPR &  & 7 & (33.3\%) & 3 & (30.0\%) & 0 & (00.0\%) & 5 & (33.3\%)& 2 & (13.3\%) & 17 & (24.3\%) \\
                \\
       \theme {Legal basis} & &   & 18 & (85.7\%) & 7 & (70.0\%) & 7 & (77.8\%) & 13 & (86.7\%) & 15 & (100.0\%) & 60 & (85.7\%)\\

                    \sub & \code Consent & & 14 & (66.7\%) & 7 & (70.0\%)& 7 & (77.8\%)& 11 & (73.3\%) & 15 & (100.0\%)& 54 & (77.1\%)\\  

                    \sub & \code legitimate interest & & 10 & (47.6\%) & 2 & (20.0\%)  & 1 & (11.1\%)& 8 & (53.3\%)& 6 & (40.0\%) & 27 & (38.6\%)  \\

                    \sub & \code Contractual obligations & & 8 & (38.1\%) & 4 & (40.0\%) & 1 & (11.1\%)  & 4 & (26.7\%)& 9 & (60.0\%)& 26 & (37.1\%) \\

                    \sub & \code Public interest & & 7 & (33.3\%) & 1 & (10.0\%)  & 0 & (00.0\%) & 5 & (33.3\%) & 6 & (40.0\%) & 19 & (27.1\%)  \\

                    \sub & \code Legal obligation & & 6 & (28.6\%) & 1 & (10.0\%)  & 0 & (00.0\%) & 5 & (33.3\%) & 5 & (33.3\%) & 17 & (24.3\%)  \\

                    \sub & \code Vital interest & & 5 & (23.8\%) & 0 & (00.0\%)  & 0 & (00.0\%) & 3 & (20.0\%) & 3 & (20.0\%) & 11 & (15.7\%)  \\

                    \sub & \code Other law & & 2 & (09.5\%) & 0 & (00.0\%)  & 0 & (00.0\%) & 0 & (00.0\%) & 4 & (26.7\%) & 6 & (08.6\%)  \\

                    \sub & \code Studies or research bodies & & 1 & (04.8\%) & 1 & (10.0\%)  & 0 & (00.0\%) & 0 & (00.0\%) & 3 & (20.0\%) & 5 & (07.1\%)  \\

                    \subend & \code The same as the GDPR & &9 & (42.9\%) &2 & (20.0\%) & 0 & (00.0\%) & 4 & (26.7\%) & 3 & (20.0\%) & 18 & (25.7\%) \\ 

                   \hline
    \end{tblr}}
    \caption{Table for codes regarding interdisciplinary challenges for data protection and understanding of data protection. Codes are unique instances, subcodes can overlap between participants. $^\dag$:\,Inductive codes. } \label{tab:challenges}
        \vspace{-2em}
\end{table*}
 


\begin{table*}\ContinuedFloat
\SetTblrInner{rowsep=0.25pt} 

    \centering
    \scriptsize

    \begin{tblrtikzabove}
        \mytreesub{sub}
        \mytreesubend{subend}
        \mytreepass{pass}
    \end{tblrtikzabove}

\SetTblrOuter{expand=\theme\code\pass\sub\subend}
    \makebox[\textwidth]{\begin{tblr}{
        colspec={Q[l,m,1cm]@{\,}Q[l,m,1cm]@{\,}Q[l,m]@{}Q[r, m]Q[l,m]Q[r, m]Q[l,m]Q[r, m]Q[l,m]Q[r, m]Q[l,m]Q[r, m]Q[l,m]Q[r, m]Q[l,m]}, 
        columns = {colsep=2pt},
        cell{1}{1,2,3} = {r=2}{m},  
        column{Z,Z[2]}={gray9,font=\bfseries},  
        hline{2} = {4-Z}{},  
        cell{theme} = {c=3}{l},
        cell{code} = {c=2}{l},
        hspan=even,
        rowsep={0.25pt},
        }
     Theme & Code & Subcode & \SetCell[c=12]{c} {Number of occurrences (percentage of n in column)}  &&&&&&&&&&& \\   
        &&& \SetCell[c=2]{c}{Asia, Oceania \\ n = 21} && \SetCell[c=2]{c}{Africa \\n = 10} && \SetCell[c=2]{c}{N. America\\ n = 9} && \SetCell[c=2]{c}{Europe \\ n = 15} && \SetCell[c=2]{c}{S. America \\n = 15} && \SetCell[c=2]{c}{Total \\n = 70}& \\ 

    \theme  \hline Children/minor data & &   & 20 & (95.2\%) & 9 & (90.0\%)  & 8 & (88.9\%) & 14 & (93.3\%) & 15 & (100\%) & 66 & (94.3\%)      \\
        \sub & \code Provision exist & & 11 & (52.4\%) & 7 & (70.0\%) & 5 & (55.6\%) & 10 & (66.7\%) & 10 & (66.7\%)&  43 & (61.4\%) \\
        
        \sub & \code High risk/vulnerable & & 7 & (33.3\%) & 5 & (50.0\%) & 0 & (00.0\%) & 3 & (20.0\%) & 3 & (20.0\%)&  18 & (25.7\%) \\
        
            \pass & \sub &Special categories &  3 & (14.3\%) & 3 & (30.0\%) & 0 & (00.0\%) & 0 & (00.0\%) & 2 & (13.3\%)&  8 & (11.4\%) \\
            \pass & \subend & Vulnerable &  2 & (09.5\%) & 2 & (20.0\%) & 0 & (00.0\%) & 3 & (20.0\%) & 0 & (00.0\%)&  7 & (10.0\%) \\
            
        \sub & \code Other regulation$^\dag$ & & 7 & (33.3\%) & 1 & (10.0\%) & 3 & (33.3\%) & 3 & (20.0\%) & 4 & (26.7\%)&  18 & (25.7\%) \\
        \sub & \code No provision$^\dag$ & & 4 & (19.1\%) & 1 & (10.0\%) & 1 & (11.1\%) & 0 & (00.0\%) & 2 & (13.4\%)&  8 & (11.4\%) \\
        \sub & \code Not sure$^\dag$ & & 2 & (09.5\%) & 0 & (00.0\%) &0 & (00.0\%) & 2 & (13.3\%) & 3 & (20.0\%)&  7 & (10.0\%) \\
        \sub & \code Progressive autonomy$^\dag$ & & 0 & (00.0\%) & 1 & (00.0\%) & 1 & (00.0\%) & 0 & (00.0\%) & 3 & (20.0\%)&  5 & (07.1\%) \\
        \subend & \code Interpretation$^\dag$ & & 2 & (09.5\%) & 0 & (00.0\%) & 1 & (11.1\%) & 0 & (00.0\%) & 1 & (06.7\%)&  4 & (05.7\%) \\
        \\
    \theme  Power asymmetry & &   & 14 & (66.7\%) & 7 & (70.0\%)  & 6 & (66.7\%) & 12 & (80.0\%) & 13 & (86.7\%) & 52 & (74.3\%)      \\
        
        \sub & \code Vulnerability & & 5 & (23.8\%) & 2 & (20.0\%) & 2 & (22.2\%) & 9 & (60.0\%) & 6 & (40.0\%)&  24 & (34.3\%) \\
        
            \pass & \sub & Contextual &  3 & (14.3\%) & 0 & (00.0\%) & 0 & (00.0\%) & 8 & (46.7\%) & 2 & (13.3\%)&  12 & (17.1\%) \\
            \pass & \sub & Characteristic &  3 & (14.3\%) & 1 & (10.0\%) & 0 & (00.0\%) & 0 & (00.0\%) & 5 & (33.3\%)&  9 & (12.9\%) \\
            \pass & \subend & Historical group &  0 & (00.0\%) & 0 & (00.0\%) & 1 & (11.1\%) & 1 & (06.7\%) & 1 & (06.7\%)&  3 & (04.3\%) \\
            
        \sub & \code Special category of data$^\dag$ & & 2 & (09.5\%) & 2 & (20.0\%) & 3 & (33.3\%) & 2 & (13.3\%) & 3 & (20.0\%)&  12 & (17.1\%) \\
        \sub & \code Employment & & 4 & (19.1\%) & 1 & (10.0\%) & 0 & (00.0\%) & 5 & (33.3\%) & 1 & (06.7\%)& 11 & (15.7\%) \\
        \sub & \code Children$^\dag$ & & 2 & (09.5\%) & 2 & (20.0\%) & 0 & (00.0\%) & 5 & (33.3\%) & 1 & (06.7\%)&  10 & (14.3\%) \\
        \sub & \code Not sure$^\dag$ & & 1 & (04.8\%) & 1 & (10.0\%) & 1 & (11.1\%) & 3 & (20.0\%) & 4 & (26.7\%)&  10 & (14.3\%) \\
        \sub & \code Not written$^\dag$ & & 5 & (23.8\%) & 1 & (10.0\%) & 0 & (00.0\%) & 0 & (00.0\%) & 2 & (13.3\%)&  8 & (11.4\%) \\
        \subend & \code Other laws$^\dag$ & & 0 & (00.0\%) & 0 & (00.0\%) & 2 & (22.2\%) & 4 & (26.7\%) & 0 & (00.0\%)&  7 & (10.0\%) \\
        \\
     \theme Data breaches & & & 17 & (80.1\%) & 9 & (90.0\%) & 6 & (66.7\%) & 13 & (86.7\%) & 13 & (86.7\%) & 58 & (82.9\%)\\
                    
                    \sub & \code Notification & &14 & (66.7\%) & 8 & (80.0\%) & 4 & (44.4\%) & 11 & (73.3\%) & 13 & (86.7\%)&  50 & (71.4\%) \\
                    
                    \pass & \sub & Authority & 13 & (61.9\%) & 7 & (70.0\%) & 3 & (33.3\%) & 9 & (60.0\%) & 12 & (80.0\%)&  44 & (62.9\%)\\
                    
                    \pass & \subend & Data subject & 9 & (42.9\%) & 5 & (50.\%) & 3 & (33.3\%) & 7 & (46.7\%) & 12 & (80.0\%) &   36 & (51.4\%)  \\
                    
                    \sub & \code Incident Response &  & 12 & (57.1\%)& 9 & (90.0\%)& 5 & (55.5\%)& 8 & (53.3\%) &9 & (60.0\%)&  43 & (61.4\%)\\
                    
                    \pass & \sub & Containment & 9 & (42.3\%) & 7 & (70.0\%) & 2 & (22.2\%) & 4 & (26.7\%) & 5 & (33.3\%) &  27 & (38.6\%) \\

                    \pass & \subend & Internal Notification & 3 & (14.3\%) & 1 & (10.0\%) & 2 & (22.2\%) & 0 & (00.0\%) & 1 & (06.7\%) &  7 & (10.0\%)\\
                
                    \sub & \code Type of Incident &  & 14 & (66.7\%)& 7 & (70.0\%) & 2 & (22.2\%)& 7 & (46.7\%) & 7 & (46.7\%)&  36 & (51.4\%) \\
                    
                    \sub & \code Risk of data & & 8 & (38.1\%)& 4 & (40.0\%) & 3 & (33.3\%) & 5 & (33.3\%) & 10 & (66.7\%)&  30 & (42.9\%) \\
                    
                    \pass & \subend & {Type of personal data} & 5 & (23.8\%) & 2 & (20.0\%) & 2 & (22.2\%) & 2 & (13.3\%) & 7 & (46.7\%) &  18 & (25.7\%)  \\

                    \sub & \code Time & &5 & (23.8\%)& 2 & (20.0\%) & 0 & (00.0\%) & 5 & (33.3\%)& 3 & (20.0\%)&  15 & (21.4\%) \\

                    \sub & \code Consequences (reputation) &  &5 & (23.8\%)& 2 & (20.0\%) & 1 & (11.1\%) & 1 & (06.7\%)& 1 & (06.7\%)&  10 & (14.3\%)  \\
                    
                    \pass & \subend & Public relations  & 4 & (19.1\%) & 0 & (00.0\%)& 1 & (11.1\%) & 1 & (06.7\%) & 1 & (06.7\%) &  7 & (10.0\%)  \\
                    
                    \subend & \code {Interdisciplinary team \\(legal and tech)\textsuperscript{\dag} }&   &2 & (09.5\%)& 0 & (00.0\%)& 3 & (33.3\%) & 0 & (00.0\%)& 1 & (06.7\%) &  6 & (08.6\%)\\
            \\
            \hline
    \end{tblr}}
    \caption{(cont.) $^\dag$:\,Inductive codes.} \label{tab:challenges2}
\end{table*}



\begin{table*}\ContinuedFloat
\SetTblrInner{rowsep=0.25pt} 

    \centering
    \scriptsize

    \begin{tblrtikzabove}
        \mytreesub{sub}
        \mytreesubend{subend}
        \mytreepass{pass}
    \end{tblrtikzabove}

\SetTblrOuter{expand=\theme\code\pass\sub\subend}
    \makebox[\textwidth]{\begin{tblr}{
        colspec={Q[l,m,1cm]@{\,}Q[l,m,1cm]@{\,}Q[l,m]@{}Q[r, m]Q[l,m]Q[r, m]Q[l,m]Q[r, m]Q[l,m]Q[r, m]Q[l,m]Q[r, m]Q[l,m]Q[r, m]Q[l,m]}, 
        columns = {colsep=2pt},
        cell{1}{1,2,3} = {r=2}{m},  
        column{Z,Z[2]}={gray9,font=\bfseries},  
        hline{2} = {4-Z}{},  
        cell{theme} = {c=3}{l},
        cell{code} = {c=2}{l},
        hspan=even,
        rowsep={0.25pt},
        }
     Theme & Code & Subcode & \SetCell[c=12]{c} {Number of occurrences (percentage of n in column)}  &&&&&&&&&&& \\   
        &&& \SetCell[c=2]{c}{Asia, Oceania \\ n = 21} && \SetCell[c=2]{c}{Africa \\n = 10} && \SetCell[c=2]{c}{N. America\\ n = 9} && \SetCell[c=2]{c}{Europe \\ n = 15} && \SetCell[c=2]{c}{S. America \\n = 15} && \SetCell[c=2]{c}{Total \\n = 70}& \\ 

    \theme  \hline Children/minor data & &   & 20 & (95.2\%) & 9 & (90.0\%)  & 8 & (88.9\%) & 14 & (93.3\%) & 15 & (100\%) & 66 & (94.3\%)      \\
        \sub & \code Provision exist & & 11 & (52.4\%) & 7 & (70.0\%) & 5 & (55.6\%) & 10 & (66.7\%) & 10 & (66.7\%)&  43 & (61.4\%) \\
        
        \sub & \code High risk/vulnerable & & 7 & (33.3\%) & 5 & (50.0\%) & 0 & (00.0\%) & 3 & (20.0\%) & 3 & (20.0\%)&  18 & (25.7\%) \\
        
            \pass & \sub &Special categories &  3 & (14.3\%) & 3 & (30.0\%) & 0 & (00.0\%) & 0 & (00.0\%) & 2 & (13.3\%)&  8 & (11.4\%) \\
            \pass & \subend & Vulnerable &  2 & (09.5\%) & 2 & (20.0\%) & 0 & (00.0\%) & 3 & (20.0\%) & 0 & (00.0\%)&  7 & (10.0\%) \\
            
        \sub & \code Other regulation $^\dag$ & & 7 & (33.3\%) & 1 & (10.0\%) & 3 & (33.3\%) & 3 & (20.0\%) & 4 & (26.7\%)&  18 & (25.7\%) \\
        \sub & \code No provision$^\dag$ & & 4 & (19.1\%) & 1 & (10.0\%) & 1 & (11.1\%) & 0 & (00.0\%) & 2 & (13.4\%)&  8 & (11.4\%) \\
        \sub & \code Not sure$^\dag$ & & 2 & (09.5\%) & 0 & (00.0\%) &0 & (00.0\%) & 2 & (13.3\%) & 3 & (20.0\%)&  7 & (10.0\%) \\
        \sub & \code Progressive autonomy$^\dag$ & & 0 & (00.0\%) & 1 & (00.0\%) & 1 & (00.0\%) & 0 & (00.0\%) & 3 & (20.0\%)&  5 & (07.1\%) \\
        \sub & \code Interpretation$^\dag$ & & 2 & (09.5\%) & 0 & (00.0\%) & 1 & (11.1\%) & 0 & (00.0\%) & 1 & (06.7\%)&  4 & (05.7\%) \\

        \subend & Resistance\textsuperscript{\dag} & & 1 & (04.8\%) & 0 & (00.0\%) & 1 & (11.1\%) & 3 & (20.0\%)& 1 & (06.7\%) &  6 & (08.6\%) \\
    \\
    \theme Transborder Data Flows  & &   & 20 & (95.2\%) & 9 & (90.0\%)  & 8 & (88.9\%) & 14 & (93.3\%) & 15 & (100\%) & 66 & (94.3\%)      \\
        \sub & Allowed$^\dag$ && 0 & (00.0\%) & 1 & (10.0\%) & 3 & (33.3\%) & 0 & (00.0\%) & 2 & (13.3\%) & 6 & (8.57\%)\\
        \sub & \code Allowed - similar to EU && 15 & (71.4\%) & 7 & (70.0\%) & 2 & (22.2\%) & 7 & (46.7\%) & 8 & (53.3\%) & 39 & (55.7\%)\\
            \pass & \sub & Adequacy decision & 9 & (42.9\%) & 4 & (40.0\%) & 0 & (00.0\%) & 4 & (26.7\%) & 6 & (40.0\%) & 23 & (32.9\%)\\
            \pass & \sub &SCC  & 7 & (33.3\%) & 2 & (20.0\%) & 1 & (11.1\%) & 4 & (26.7\%) & 5 & (33.3\%) & 19 & (27.1\%)\\
            \pass & \sub & BCR & 5 & (23.8\%) & 1 & (10.0\%) & 1 & (11.1\%) & 2 & (13.3\%) & 4 & (26.7\%) & 13 & (18.6\%)\\
            \pass & \sub & Similar/ as the GDPR  & 9 & (42.9\%) & 5 & (50.0\%) & 0 & (00.0\%) & 7 & (46.7\%) & 5 & (33.3\%) & 26 & (37.1\%)\\
            \pass & \sub & Treaty & 4 & (19.1\%) & 1 & (10.0\%) & 2 & (22.2\%) & 0 & (00.0\%) & 3 & (20.0\%) & 10 & (14.3\%)\\
            \pass & \sub & Consent (exception) & 3 & (14.3\%) & 2 & (20.0\%) & 0 & (00.0\%) & 1 & (6.67\%) & 0 & (00.0\%) & 6 & (8.57\%)\\
            \pass & \sub & Seals and certificates & 4 & (19.1\%) & 0 & (00.0\%) & 0 & (00.0\%) & 0 & (00.0\%) & 1 & (6.67\%) & 5 & (7.14\%)\\
            \pass & \subend & Certifications & 4 & (19.1\%) & 0 & (00.0\%) & 0 & (00.0\%) & 0 & (00.0\%) & 1 & (6.67\%) & 5 & (7.14\%) \\
            
        \subend & \code Other$^\dag$ && 12 & (57.1\%) & 3 & (30.0\%) & 3 & (33.3\%) & 1 & (06.7\%) & 6 & (40.0\%) & 25 & (35.7\%)\\
             & \sub & Consent (common)$^\dag$ & 6 & (28.6\%) & 1 & (10.0\%) & 1 & (11.1\%) & 0 & (00.0\%) & 3 & (20.0\%) & 11 & (15.7\%)\\
             & \sub & {Equivalent level of\\ protection$^\dag$} & 5 & (23.8\%) & 2 & (20.0\%) & 1 & (11.1\%) & 0 & (00.0\%) & 2 & (13.3\%) & 10 & (14.3\%)\\
             & \subend & {Performance of \\a Contract$^\dag$} & 4 & (19.1\%) & 1 & (10.0\%) & 3 & (33.3\%) & 0 & (00.0\%) & 0 & (00.0\%) & 8 & (11.4\%)\\
    \\
    \theme  Characteristic of data protection & &   & 12 & (57.1\%) &  5 & (45.5\%)  & 3 & (33.3\%) & 10 & (66.7\%)  & 12 & (80.0\%) & 42 & (60.0\%)      \\
        \subend & \code {Combination \\(tech, legal and procedure)} && 11 & (52.4\%) & 5 & (45.5\%) & 3 & (33.3\%) & 10 & (66.7\%) & 12 & (80.0\%) & 41 & (58.6\%)\\
\hline
\end{tblr}}
    \caption{(cont.) $^\dag$:\,Inductive codes.} \label{tab:challenges3}
\end{table*}

We did not collect the nationality of origin, but rather the country/region of expertise. Multiple participants exercised as legal experts in jurisdictions where they were not necessarily nationals. More information on the participants can be obtained at \Cref{tab:summary-participants}. We provide only high level data demographics, for anonymization purposes, to avoid re-identification. We, thus, divide our sample based on continent only for anonymization purposes. 

We did not manage to get individuals from Saudi Arabia or Azerbaijan due to no-shows. Most interviews were carried out using the University's online videoconference platform, while 7 were done in person and 2 were done over other encrypted videoconference platforms, upon the request of the interviewee. 57 interviews were conducted in English, 15 in Spanish, two in French.
All individuals except 5 consented on having their interviews recorded, and some consented on having their anonymized transcripts made available if requested by other researchers.  A minority of interviewees agreed on making their transcripts publicly available. 

Contrary to what we had planned, most subjects could not be labeled in a single category of practice; i.e., academic, activist, regulator or private practitioners. In most cases, they were labeled with at least two or more categories at the same time. 
Some specific cases could be labeled in one category (mostly academics), but due to re-identification and privacy risks, we do not provide these.

\subsection{Results from the regulatory analysis}

{\scriptsize
\begin{longtable}[c]{p{4.9cm}ccc} 
\caption{Summary of the regulatory analysis, frequency per code. $^\star$: The APPI from Japan does not directly foresee the right to revoke consent. $^\star$$^\star$: \Cref{annex:detailed-analysis-regulation} provides an explanation over this aspect.}
\label{tab:summary-regulations-short} \\

 & \multicolumn{3}{c}{Frequency} \\ \cline{2-4}
Requirement code & Exist & Partial & {\begin{tabular}{@{}l@{}}Not conceptualized the\\ same or not mentioned\end{tabular}}  \\ \hline
\endfirsthead

\multicolumn{4}{c}%
{{\footnotesize \tablename\ \thetable{}: Summary of the regulatory analysis, frequency per code (continued)}} \\
 & \multicolumn{3}{c}{Frequency} \\ \cline{2-4}
Requirement code & Exist & Partial & {\begin{tabular}{@{}l@{}}Not conceptualized the\\ same or not mentioned\end{tabular}}  \\ \hline
\endhead

\hline
\multicolumn{4}{r@{}}{{\normalsize Continued on next page}} \\
\endfoot

\hline
\endlastfoot

\multicolumn{4}{c}{\cellcolor{grey!60} Conceptualization of personal data}  \\
Identifiable  & 21 & 0 & 0  \\
Identified  & 21 & 0 & 0  \\
Legal persona  & 2 & 0 & 19  \\
Natural persona & 21 & 0 &0  \\
\hspace{1em} $\hookrightarrow$ Alive (explicit)& 12 & 0 & 9 \\
\hspace{1em} $\hookrightarrow$ Deceased& 0 & 7 & 14\\
Special categories of data (Sensitive) & 19 & 2 & 0 \\[0.6em]

\multicolumn{4}{c}{\cellcolor{grey!60}Data subject right}  \\
To be informed  & 21 & 0 & 0  \\
Access  & 21 & 0 & 0  \\
\hspace{1em} $\hookrightarrow$ Free of charge& 8 & 4 & 9 \\
Rectification  & 21 & 0 & 0  \\
Erasure ($\neq$ to be forgotten) & 21 & 0 &0  \\
To Restrict Processing & 14 & 0 & 7 \\ 
Portability & 8 &  1 & 12 \\
Object/Opt-out & 10 & 4 & 7 \\ 
\hspace{1em} $\hookrightarrow$ Object direct marketing & 10 & n/a & n/a\\ 
Revoke consent & 20 & 0 & 1\rlap{$^\star$} \\
Not to be subject to automated decision-making & 11 & 0 & 10 \\
Provide ID & 16 & 1 & 4 \\ 
Complains to authority & 21 & 0 & 0 \\ 
Time frame to provide right & 19 & 0 & 2 \\
All DSR free of charge & 6 & 5 & 10 \\[0.6em] 

\multicolumn{4}{c}{\cellcolor{grey!60}Legal bases}  \\
Consent  & 21 & 0 & 0  \\
\hspace{1em} $\hookrightarrow$ As the norm& 7 & 0 & 14 \\
\hspace{1em} $\hookrightarrow$ Free& 21 & 0 & 0 \\
\hspace{1em} $\hookrightarrow$ Informed& 20 & 0 & 1 \\
\hspace{1em} $\hookrightarrow$ Explicit& 20 & 0 & 1 \\

Legitimate interest  & 12 & 0 & 9  \\
Contractual obligations & 17 & 0 & 4\\
Legal obligation & 21 & 0 & 0 \\
Vital interest & 19 & 0 & 2 \\
Public task &  18 & 0 & 3 \\ 
Other legal bases for SP & 19 & 0 & 2 \\[0.6em] 

\multicolumn{4}{c}{\cellcolor{grey!60} Actors or roles mentioned}  \\
Data subject (individual) & 21 & 0 & 0 \\
Controller & 19 & 2\rlap{$^\star$$^\star$} & 0 \\
Processor & 18 & 0 & 3 \\
Data Protection Authority & 21 & 0 & 0 \\
Data protection officer & 17 & 3 & 1 \\[0.6em]

\multicolumn{4}{c}{\cellcolor{grey!60} Children}\\
Identified in the DP law & 12 & 0 & 9 \\[0.6em]

\multicolumn{4}{c}{\cellcolor{grey!60}Transborder flows}  \\*  
Allowed per conditions  & 21 & 0 & 0  \\

SCC  & 10 & 0 & 11  \\
BCR & 11 & 0 & 10\\
Consent (common) & 10 & 1 & 10 \\
Consent (derogation) & 7 & 1 & 13 \\
Adequacy list &  9 & 2 & 10 \\[0.6em]

\multicolumn{4}{c}{\cellcolor{grey!60} Security requirements for RDPR} \\
Security requirements & 21 & 0 & 0 \\
Data breach definition in DP law & 13 & 0 & 8 \\
Notification & 20 & 0 & 1 \\ 
\hspace{1em} $\hookrightarrow$ Authority& 16 & 0 & 5 \\
\hspace{1em} $\hookrightarrow$ Data subject& 18 & 0 & 3\\
\hspace{1em} $\hookrightarrow$ Time frames & 17 & 0 & 4 \\
Fine & 20 & 0 & 1 \\
Training & 15 & 0 & 6 \\[0.6em]

\multicolumn{4}{c}{\cellcolor{grey!60} Documentation}\\
Privacy policy & 21 & 0 & 0 \\
Privacy notice & 21 & 0 & 0 \\
RPA & 12 & 0 & 9 \\
(D)PIA & 13 & 2 & 6 \\
Registration with authorities & 8 & 1 & 12 \\[0.6em]

\multicolumn{4}{c}{\cellcolor{grey!60}Privacy notice information}  \\
Identity and contact details of controller & 18 & 0 & 3  \\
DPO contact details  & 4 & 0 & 17  \\
Purpose and legal basis  & 21 & 0 & 0  \\
Legitimate interest explanation & 3 & 0 & 18  \\
Recipient or categories of recipients of personal data & 13 & 0 & 8 \\ 
Retention time & 6 & 0 & 15 \\
Explain legal or contractual obligations & 5 & 0 & 16\\
International transfers information & 7 &  0 & 14 \\
Explanation of DSR & 17 &  0 & 4 \\
\hspace{1em} $\hookrightarrow$ General provision (or) & 10 & n/a & n/a\\ 
\hspace{1em} $\hookrightarrow$ Detailed provision per right (or) & 3 & n/a & n/a\\ 
\hspace{1em} $\hookrightarrow$ Specific rights (or) & 4 & n/a & n/a\\ 

\end{longtable}
}

As part of the triangulation efforts, we keep track where our codes appear in the regulation. Most countries in our sample have a pan-regulation on personal data protection; however, some have multiple legal documents on the subject. The only exception was the United States, which in comparison to other common law countries such as Canada or Australia that have a federal data protection law, the USA does not. Hence, it is not included in our triangulation effort. Similarly, most regulations referenced other legal documents (such as cybersecurity laws). Given the scope of this project, we focused primarily on data protection regulation, except when specialized literature pointed to other regulations and emphasized their importance. We discuss this in further detail in \Cref{sec:threats}.

To provide the summary of the regulation notation, we present our codes: exist, partial, and not conceptually the same. We define `exist’ if the regulation presents the code with the same or very similar requirements. If the requirements had similarities but were not exactly the same, we coded as `partial'. And if there were little to no similarities, we coded as `not conceptually similar’. Our public repository presents more details on this coding\footnote{Our data availability statement is available at \Cref{sec:data-availability-statement} or at \url{https://doi.org/10.5281/zenodo.15166882}}. 

All of these documents are summarized in \Cref{tab:summary-regulations-short}. A more detailed version is available in \Cref{tab:regulation}, and a fully detailed version, with each code corresponding to the corresponding section of a regulation, is available in our public repository. Furthermore, the table in the public repository points out which regulation(s) were analyzed, with the corresponding article or section of the regulation. Finally, \Cref{tab:sources-sca-regulation} lists the specialized literature used to support our analysis.

\subsection{SQ1: Which data protection requirements are common across regulations, and how are they conceptualized?}

If fewer than 66\% of the regulations, which implies 14 regulations, did not impose such requirements, we would  not include them in our conceptualization. Quotes translated from Spanish or French were translated by the authors, with the original quotes in the \Cref{app:original-quotes}. 

\begin{longtblr}
[
  caption = {List of sources used, per country, to support the SCA for the main regulatory data protection analysis, including the main data protection regulation analyzed.},
  label= {tab:sources-sca-regulation},
]{
  colspec={lXX},
  rowhead=1,
  cells={font=\scriptsize}
}
Country    & Sources used to support coding                                                                                  & Main data protection regulation \\
\hline
Argentina  & \cite{Ley25326,IAPPComparison,compliancelatam2025guia,argentina_datoPersonales,DataGuidanceArgentina,DataGuidanceDSRArgentina,aaip2023derechos,anchorena2025proyectoley,fernandez2017aaip,ripd2023guia,aaip2024transferencias,aaip_datospersonales,dlapiper2025argentina} &  Ley 25.326 \cite{Ley25326} and different resolutions, such as Disposición 18/2015 \cite{Disposicion182015}  \\
Australia  &  \cite{walters2019,IAPPComparison,DataGuidanceAussia,DataGuidanceDSRAussie,watts2019privacy,oaic2022apps,oaic2025app8,Girot2022Australia,dataguidance_gdpr_australia,ABLI2020CrossBorderPersonalDataFlowsAsia,Paulger2022BalancingAccountabilityPrivacyAPAC}  & The Privacy Act \cite{PrivacyActAussi}    \\
Azerbaijan &  \cite{haciyeva2021personal,DataGuidanceAzer,DataGuidanceDSRAzer,dlapiper2022azerbaijan_law,AzerbaijanDPLaw}  &   Law No. 27 of 2022 on Personal Data Protection \cite{AzerbaijanDPLaw} \\
Brazil & \cite{doneda2013data,machado2024protection,anpd2021resolucao1,LGPD,peretti2024anpd,IAPPComparison,compliancelatam2025guia,codingrights2016brazil,DataGuidanceBrazil,DataGuidanceDSRBra,fico2024minimization} & Lei Geral de Proteção de Dados Pessoais (LGPD) \cite{LGPD} \\
Canada & \cite{gratton2017droit,scassa2019data,opc2018tentips,opc2009crossborder,IAPPComparison,dataguidance_gdpr_pipeda,shibleyrighton2025canada,cba2025childrens,jaar2008canadian} & Personal Information Protection and Electronic Documents Act (PIPEDA) \cite{PIPEDA}\\
Chile & \cite{compliancelatam2025guia,poderjudicialtv2020habeas,DataGuidanceDSRChile,Ortiz_Mesías_Viollier_2021} & Ley 21.719 \cite{Ley21719} \\
China & \cite{IAPP2021PIPLvsGDPR,setiawati2020optimizing,hornuf2023data,Zhang2022GDPRVersusPIPL,GBT35273_2020PersonalInformationSecuritySpecification,NPC2020ProtectionOfMinorsLaw,DLAPiper2026ChinaDataProtection,creemers2022china,OneTrustDataGuidanceGDPRvPIPL,DataGuidanceDSRPChina,DataGuidanceChina,IAPPComparison,Girot2020TransferringPersonalDataAsia,ABLI2020CrossBorderPersonalDataFlowsAsia} & Personal Information Protection Law of the People's Republic of China \cite{PIPL} \\
Colombia & \cite{cano2019alcance,lloredacamacho2021colombia,sic2025abc,cabezas2023tratamiento,gomez2020derecho,balanta2020guia,DataGuidanceColombia,DataGuidanceDSRPColombia} & Ley 1581 de 2012 \cite{Ley1581} and \cite{colombia2013decreto1377}\\ 
Egypt & \cite{clydeco2020egypt,andersen2025law151,pdpc2026consent,el2026report,daigle2021data,DataGuidanceEgypt} & The Data Protection Law, Resolution No.151 [non official translation to English] \cite{andersen2025law151}]\\
India & \cite{sundara2023protecting,naithani2025analysis,santana2023data,DataGuidanceIndia,DataGuidanceaDSRIndia,garimella2024india,Girot2020TransferringPersonalDataAsia,ABLI2020CrossBorderPersonalDataFlowsAsia} & Digital Personal Data Protection Act 2023 \cite{DPDA} and The Information Technology (reasonable Security Practices And Procedures And Sensitive Personal Data Or Information) Rules, 2011 \cite{india2011itrules} \\
Indonesia & \cite{penasthika2024indonesia,abnr2025indonesia,taufiq2025legal,maleno2024comparative,ABLI2020CrossBorderPersonalDataFlowsAsia,DataGuidanceDSRIndonesia,DataGuidanceIndonesia} & Law No. 27 of 2022 on Personal Data Protection\cite{PDPLaw}\\

Japan & \cite{japan2003cabinetorder507,JapanAdequacy,Sato2024,adams2009japanese,DataGuidanceJapan,DataGuidanceJapan2,Girot2020TransferringPersonalDataAsia,ABLI2020CrossBorderPersonalDataFlowsAsia,Paulger2022BalancingAccountabilityPrivacyAPAC,dataguidance_gdpr_appi} & Act on the Protection of Personal Information \cite{APPI}\\
Mexico & \cite{garrigues2025mexico,compliancelatam2025guia,DataGuidanceDSRMex,DataGuidanceMexico,dataguidance_gdpr_mexico}& Ley {Federal} de {Protección} de {Datos} {Personales} en {Posesión} de los {Particulares} ({LFPDPPP}) \cite{mexico2025lfpdppp} \\
Saudi Arabia & \cite{sdaia_about,alzahrani2024overview,dataguidance2025gdpr_pdpl,almanea2025pdpl,DataGuidanceSaudi,DataGuidanceDSRSaudi}& The Personal Data Protection Law - Royal Decree No. M/19 of 9/2/1443 H (PDPL) \cite{PDPLSaudi} \\
Singapore & \cite{setiawati2020optimizing,pdpc2023overview,pdpc2026guidelines,dataguidance2022gdpr_pdpa,lim2017gdpr,DataGuidanceDSRSingapore,DataGuidanceSingapore,IAPPComparison,ABLI2020CrossBorderPersonalDataFlowsAsia,Girot2020TransferringPersonalDataAsia}& Personal Data Protection Act 2012 (2020 revised edition) \cite{singapore2021pdpr} and guidelines provided by the Authority \\
South Africa & \cite{abdulrauf2019data,nel2017popia,jones2022popia,dataguidance_gdpr_popia,DataGuidanceDSRSouthAfrica,DataGuidanceSouthAfrica,IAPPComparison}& Protection of Personal Information Act (POPIA) \cite{POPIA} \\
South Korea & \cite{setiawati2020optimizing,pipc2024guidelines,kim2024automated,JOO2023101805,dataguidance2023gdpr_pipa,ComprehensiveGuidetoSouthKoreasPIPA,korea2025pipaDecree,DataGuidanceDSRKorea,DataGuidanceKorea,Girot2020TransferringPersonalDataAsia,ABLI2020CrossBorderPersonalDataFlowsAsia,IAPPComparison} &  Personal Information Protection Act \cite{PIPA} \\
Tanzania & \cite{malekela2025privacy,boshe2016data,DataGuidanceTanzania,DataGuidanceTanzania2,MalekelaJurisic2025TanzaniaPDPDArbitration} &  The Personal Data Protection Act \cite{PDPATanzania}\\
Turkey & \cite{kvkk_datasub_rights,kvkk2018communique6638,kvkk2017bylaw6636,kvkk2019decision10,linklaters_turkey,IAPPComparison,DataGuidanceDSRPTurkey,DataGuidanceTurkey} & Law on the Protection of Personal Data No. 6698 (LPPD) \cite{LPPD}\\
United Kingdom & \cite{ICO2026LawfulBasis,ICO2026InternationalTransfers,ICO2023IndividualRights,DataGuidanceUK,DataGuidanceDSRUK} & UK General Data Protection Regulation (UK GDPR) \cite{UKGDPR}\\ 
\hline
\end{longtblr} \label{tab:sources-regulation}
 
\paragraph{Personal data} 

We found that multiple countries indicate that any information that can lead to identification (usually phrased as \textit{`Identified'} or \textit{`Identifiable'}) is conceptualized as personal data. 51 participants (72.7\%) legally defined personal data as \textit{`Identifiable'}, compared to 39 participants (55.7\%) who described it as \textit{`Identified'}. From these two codes, co-occurrence occurred 38 times (97.4\%) .

Our regulatory analysis throws a similar conclusion, with all 21 regulations including these two codes in their framework when conceptualizing personal data (or it has been interpreted as such). We triangulated this data with previous literature, as we were not expecting this high-level of saturation, which confirmed the conclusion. For example the Colombian Data Protection \cite{Ley1581} law says `determinadas o determinables' which roughly translates to `identified or identifiable'. Participants agreed that this definition seemed standardized worldwide. ``\textit{Personal data is, I don't remember the exact definition off the top of my head, but it's any data that identifies or makes a person identifiable. It's essentially the same concept and doesn't differ much from what we see in other parts of the world''} (P 49)$^\ddagger$\footnote{Quotes marked with $^\ddagger$ have been translated from Spanish by the lead author.}. 

In addition, personal data generally refers to data relating to a living natural person. Although legal experts did not mention this code as much as others (n = 5, representing a 7.1\% under \textit{`Alive'}), twelve jurisdictions specifically indicate that personal data is of an \textit{`Alive'} \textit{`Natural Person'}. This implies that, when a natural person dies, their data is no longer classified as personal data, which can have different privacy requirements and implications. Although privacy expectations still apply after death, personal data management changes and therefore systems could benefit from a new classification \cite{edwards2013protecting}. 

Nevertheless we found some exceptions to these findings. Firstly Azerbaijan, Canada, Chile, China, India, Saudi Arabia and Singapore conceptualization of personal data (partially) includes a \textit{`Deceased'} \textit{`Natural persona'}. In this sense, there are still some RDPRs that applies to a deceased person's personal data, although not fully and not all the regulation. Hence, we classify this situation as partial. For example, a deceased individual may still have some data subject rights (DSRs) under some of these frameworks, such as Chile. One participant clarifies  \textit{``It's more like a right to nominate somebody who will be exercising your rights upon you being, you know, losing your life or otherwise being incapacitated''} (P 15). However these countries do not treat personal data from an alive natural person exactly the same as a deceased person.  

Secondly, South Africa's regulation (Section 1 of PoPIA \cite{POPIA}) and Argentina's law (Art. 2 point 1 of \cite{Ley25326}) include legal (\textit{`Non-natural'}) personas. This implies that personal data protection includes not only humans but also \textit{`Legal persona'}(s). As one participant explains, \textit{``It makes the South African law, but weird, because I think we're one of the only laws that provides data protection to an entity, a legal entity, which is kind of weird. So you can protect your business''} (P 5).

\begin{mdframed}
    Finding \findingsnumber: Personal data is conceptualized as data about an identified or identifiable natural living person. 
\end{mdframed}

Most regulations would distinguish between different types of personal data based on their risks. For example, there are \textit{`special categories'} (SP) of personal data. In our interviews, we coded SPs in 22 (31.4\%) of the interviews, as shown in \Cref{tab:summary-regulations-short}. The interviewees explained that SPs would constitute data categories that pose higher risks to subjects and therefore have different and more stringent processing requirements.

From a regulatory perspective, in 19 jurisdiction there is a clear and evident conceptualization that differentiates between \textit{SPs} and `normal' personal data, as shown in \Cref{tab:summary-regulations-short} . This distinction is widespread, with the exception of Singapore that does not make a distinction, and India\footnote{Article 3 of The Information Technology (Reasonable Security Practices and Procedures and Sensitive Personal Data or Information) Rules from 2011 (SPDI \cite{SPDI}). \textit{`SPs'} it is not, however, specified in the Digital Personal Data Protection Act from 2023 \cite{DPDA}.  The different conceptualization still seems to exists in practice, as reported by the literature \cite{naithani2025analysis}. Art. 9 makes differences for processing personal data from children or `a person with disability'  which can be interpreted as conceptualization of sensitive personal data, although this may be debatable \cite{DPDA}. Furthermore, in Art. 10 (1) (a) it makes references to the sensitivity of personal data being processed as a factor to define data fiduciaries as a significant data fiduciary.}. One participant explains \textit{``The PDPA does not have, define a category of data called sensitive personal data. But in the PDPC's decisions, it is mentioned. So it is mentioned through the regulatory decisions, but not expressly mentioned in the Act itself''} (P 22).

\begin{mdframed}
    Finding \findingsnumber: There are different categories of personal data. Special categories (or sensitive personal) personal data is usually identified as data that poses a higher risk to individuals' freedom or dignity, and therefore demands for a different data management. This different data management lifecycle imposes different requirements on the processing of data. 
\end{mdframed}

\paragraph{Actors} 

As data protection regulations relate to living natural persons, one code present across all regulations, albeit with different nomenclature, is \textit{`Data subject'}, as seen in \Cref{tab:summary-regulations-short}. It was also mentioned in 30 (46.9\%) interviews and in 21 regulations. 

As RDPRs impose requirements on organizations, 48 (68.6\%) interviewees recognized that the regulations discuss and acknowledge \textit{`controllers'}. For our participants, \textit{`Controllers'} were, at a minimum, the organization that defined the reason for processing personal data and its' management. When triangulating the information with the regulations, 19 regulations discuss controllers broadly, even if they have different names. Australia and Canada are the exception and is partially similar to our code, per the revision of specialized literature shown in \Cref{tab:sources-regulation}. 

Overall, 43 participants (61.4\%) mentioned \textit{`Controller'}, and 18 regulations distinguish between \textit{`Controller'} and \textit{`Processor'}. According to the GDPR conceptualization per DQA, controllers are distinct from processors \cite{GDPR}. Controllers define personal data lifecycle management, while processors `process' the data according to the controller's directions. Both have different responsibilities and duties. Three regulations (Argentina, Australia, and Canada) do not conceptualize or have interpretations for the processor. For example, even though Japan does not have a concept directly, it has been interpreted that these different actors exist, as discussed in the EU adequacy decision in paragraph 35 \cite{JapanAdequacy}. Such is not the case for the regulations we revised for the first three countries mentioned. 

Although the duties and responsibilities of these two roles may not be the same, which we discuss in the next question, most regulations recognize this difference.
One participant from a country in Asia, that has a heavily influenced regulation  by the GDPR indicates that the \textit{``processor-controller difference is totally the same [to the GDPR]''} (P 30). Other regulations that are not heavily inspired by the GDPR can have different nomenclature and requirements but still make the distinction through guidelines, amendments, or other legal documents. For example, one participant explained, \textit{``So, data intermediary refers specifically to processors [...] but as we go through the legislation, it's usually clear what is being referred to. And Singapore recognized the lack of clarity, sometimes in not adopting similar terms to the GDPR. So, in some of its advisory guidelines and guides, it started using the term controller and processor''} (P 22). Thus, the distinction between the entity that manages personal data (the \textit{`controller'}) and the entity that processes data subject to the controller (the \textit{`processor'}) persists across regulations.

Another commonly mentioned actor across our interviews was \textit{`Data Protection Authorities'}, with 33 participants (47.1\%) discussing it. Similarly, all regulations stipulate and mandate data protection authorities. \textit{``There are authorities, of course, so data protection authorities, which enforce the law''} (P  38), while also providing guidance and other legal artifacts. \textit{``The authority has been very active, I would say, in engaging the public, in developing guidance notes. So there's a lot of guidance notes''} (P 9). They usually have investigative powers, specific mandates to enforce the law, the authority to receive and proceed with complaints, impose fines, among other powers. Some of these authorities are still in the process of being constituted; \textit{``Unfortunately, we are still not implementing it correctly because we are missing the Personal Data Protection Authority''} (P 19).

Finally, another important actor we identified is the \textit{`Data protection officer'}. Although this code was mentioned less frequentlym with 13 participants (18.6\%), when triangulating the information with the regulations it appears as an important requirements;  more than two-thirds of the regulations (17) specify the role of data protection officer within organizations. In 3 regulations, it is considered a good practice or must be appointed under specific circumstances (which we code as partial) and only Azerbaijan's regulation does not include it. So although the name may differ, the role objective remains similar\footnote{ Their roles is to be in charge of compliance within an organization.}. As one participant said, \textit{``Then we don't have a role for the data protection officer.
We have an information officer, which is the ultimate accountable person, executive, as in executing, right? That is responsible for compliance with the legislation''} (P 5). 

\begin{mdframed}
    Finding \findingsnumber: All regulations identify data subjects and data protection authorities as actors. To a lesser extent, 18 regulations distinguish between controllers and processors. In these cases, the controller defines the personal data lifecycle, while the processor processes the data on the controller's behalf. Data protection officers are also a requirements in 17 regulations, and play vital roles in organizations' compliance.
\end{mdframed}

\paragraph{Legal basis} 
Although the terminology vary across regulations, most countries specify legal bases or purposes for data processing. Consequently, only some legal bases are common across regulations.

Consent is the only legal basis present in all 21 of the regulations examined. It is typically defined as informed, freely given, specific to a purpose, and unambiguous, consistent with our codebook \cite{GDPR}. Additionally, 54 participants (77.1\%) identified \textit{`consent'} as a legal basis. The frequent mention of consent highlights its widespread acceptance across jurisdictions and discussion across jurisdictions. As one interviewee noted: \textit{``The popular one, currently a lot of people use consent. Yeah, it's easy, and it's easier. And also like, they can guarantee that, OK, I asked your consent and I can process your data. And also you have the right to withdraw it. It's clearer and guarantees transparency''} (P 10).

Furthermore, consent should be presented in clear, understandable language, so that individuals can distinguish what they are consenting to. This relates to the \textit{`Right to be Informed'} and \textit{`Privacy Notices'}, which we discuss later, as both are also present in all regulations. Regarding the clarity and comprehensibility of consent, one participant explained that \textit{`how this information is transmitted, so it should be explained to the person in a way that is clearly understandable''} (P 65). This conceptualization establishes requirements for gathering consent and how to satisfy them. It is more than a `tick-box': \textit{``So that's why you have cases of the checkbox content, right? Because they [software engineers] know like OK, content is checkbox and that's it. So lots of training and lots of conversation needs to happen between teams [legal and IT]. ''} (P 19). As another participant explained, it is a textured requirement: \textit{So they [software engineers] want yes, no, but we need to problematize it to put the landscape at stake, and they want yes and no answers. And for us, it's more like it depends. So this was very interesting. And they do not deal well with open concept, open-textured concepts''} (P 33).

In some jurisdictions, such as Canada, Argentina, and Colombia, consent serves as the common legal basis, as reflected in the sub-code \textit{`consent'}, \textit{`as the norm'} in \Cref{tab:summary-regulations-short}. \textit{``It is rooted in the concept of consent, and, unfortunately, other legal grounds are either not incorporated at all or are incorporated in a way that requires a certain degree of interpretation''} (P 54)$^\ddagger$. This means that all other lawful processing grounds are considered derogations and should only be used when there is a very specific reason not to rely on consent.

\begin{mdframed}
    Finding \findingsnumber: Consent conceptualizations requires to be gathered in an \textit{`free', `informed', `specific'}, and \textit{`unambiguous'}. \textit{`Free'} mean that the data subject was not coerced or influenced to accept the terms and provide personal data. \textit{`Informed'} means that the data subject has sufficient information to make an informed decision and understands how the data will be processed and the risks involved. \textit{`Specific'} means that consent is given for a particular purpose and is not open-ended. \textit{`Unambiguous'} means the data subject has clearly given consent to that data processing.
\end{mdframed}

There are other common a legal bases for processing personal data in data protection regulations. \textit{`Contractual obligations'} was mentioned by 26 participants (37.1\%) and in in 17 regulations as a legal basis. This legal basis refers to the processing of data necessary to fulfill contractual obligations between parties, which the data subject is part. \textit{`Legal obligation'}, is mentioned in all regulations and by 17 (24.3\%) of participants, and refers to obligations mandated by legal provisions that the controller or processor must follow. \textit{`Vital interest'}, defined in 19 regulations, involves processing data to protect the data subject's life. And finally, the \textit{`public interest'}, mentioned in 17 regulations and by 19 (27.1\%) participants, refers to the processing of personal data in the public interest, as specified by a legal instrument or public authority.

\begin{mdframed}
    Finding \findingsnumber: Other legal bases commonly found across regulations for processing data include public task, fulfillment of a contract, legal obligation, and vital interest.
\end{mdframed}

\paragraph{Data subject rights}

The following data subject rights were discussed by participants: \textit{`Access'} to information was mentioned by 35 (50\%) participants, \textit{`Rectification'} by 35 (50\%) participants,  \textit{`Erasure'} by 31 (44.3\%) participants, \textit{`To be Informed'} by 20 (28.6\%) participants, the right \textit{`To object'} by 14 (20\%) participants, and \textit{`Portability'} by 17 (24.3\%) participants. 

Although not all participants mentioned every right, the identified ones are consistent across jurisdictions when triangulated with the regulations in our sample (as shown in \Cref{tab:summary-regulations-short} and with previous work \cite{lim2025navigating}). Specifically, the rights \textit{`To be Informed'}, \textit{`Access'}, \textit{`Rectification'}, and \textit{`Erasure'} are present in 21 regulations. The right \textit{`To Restrict Processing'} is found in 14 regulations and discussed by 11 (15.7\%) participants, while the right of \textit{`Object/Opt-out'} is explicitly recognized in 9 regulations and discussed by 14 (20\%) participants. Interviewees emphasized that the right to be informed extends beyond merely providing information to data subjects. Rather, it requires that such information is presented in a clear and understandable manner. As one participant explained, \textit{``Right [to] information in simple terms is provided to the user, and that makes it understand it, that the user understands what's going on''} (P 53). Another widespread right, present in 21 regulations respectively, is the right to file \textit{`Complains to authority'}.

Regarding the right to \textit{`Erasure'}, the conceptualization in this study makes a distinction from the so-called \textit{`Right to be Forgotten'} under the GDPR. One participant emphasized, \textit{``For example, Article 17 on the right to be forgotten, that’s a load of nonsense that makes no sense at all in Latin America. And it’s not as if Latin America doesn’t recognize the right to have information deleted'' }(P 49)$^\ddagger$. A more nuanced analysis of the distinction between the right to be forgotten and the right to erasure is provided in the following research question. In this context, the common conceptualization of the right to erasure, as reported by participants, aligns with deleting data that is no longer necessary for its purposes, or specific circumstances described by the regulation \cite{BUNN2015336}.

Another data subject right identified is the right  \textit{`To Revoke Consent}. This right was referenced less frequently by participants, with 13 (18.6\%) mentioning it. When analyzed alongside the regulations, this right appears in 20 of them, as shown in \Cref{tab:summary-regulations-short}. As participants discussed in the interviews, consent must be freely given, which implies that it is also revocable. However and Japan's regulation  \cite{APPI}  does not conceptualize this right in the same manner as defined in the codebook \cite{JapanAdequacy,Hayashi2024,DataGuidanceJapan,DataGuidanceJapan2}. For example, as one participant noted, \textit{``The European Parliament resolution in December 2018, again, on the Japanese adequacy report pointed out there's no mechanism if the data subject withdraws his or her consent''} (P 20) which is in line with what the adequacy decision states \cite{JapanAdequacy}.

The right \textit{`To Restrict Processing'} is barely over our defined threshold, being fully defined by 14 regulations and discussed by 11 (15.7\%) participants. As such, this right can be considered in this article as common; however, in a limited scope. Additional elements identified through regulatory analysis and interviews include data subjects' right to file \textit{`Complain to authority'}. This right is present in all analyzed regulations.

\begin{mdframed}
    Finding \findingsnumber: The rights to access, rectification, erasure, restriction of protections, as well as the right to revoke consent are common across regulations. The right to be informed is also present in all regulations and is implemented through various documents that must inform data subjects about the processing of their personal data. Data subjects also have the right to file complaints with the data protection authority. However, the concept of erasure varies by country, so it is important to understand the national scope and details.
\end{mdframed}

Regarding possible charges for data subjects' rights, 9 regulations  require the right to \textit{`Access'} to be \textit{`Free of charge'} without limitations\footnote{At the moment of writing, the GDPR Digital Omnibus is under discussion.}. 4 jurisdictions allow this right to be exercised free of charge, but with limitations such as the number of requests permitted within a defined time frame - as for example Chile \cite{Ley21719}. 9 jurisdictions do not require this right to be free of charge. This finding raises the question of whether a data subject's right can be considered a true right if it is subject to monetary constraints. We leave this question open for future research and discuss in more detail this topic in the next question. 

\paragraph{Security and data breach response}

Security (coded as `\textit{data breaches'}) emerged as a frequent concern among participants (58 participants, 82.9\%). Triangulating the legal matrix indicates that all 21 regulations mandate \textit{security requirements}. Historically, data protection and security have been closely linked, as security safeguards personal data from threats. Security is a core principle in the OECD privacy guidelines \cite{OECD_reportGuidelines}, CBPR \cite{CPBR,APECCBPR}, and GDPR \cite{GDPR}. It is also required for RDPRs, and several countries have enacted complementary cybersecurity laws\footnote{These laws were beyond the scope of this research; therefore, personal data security requirements may be more detailed in those regulations.}. As one participant stated: \textit{``You can't achieve data compliance without data governance, and vice versa; you can't ensure privacy without cybersecurity''} (P 68)$^\ddagger$.

Consistent with previous literature, lawyers emphasized that RDPRs cannot be satisfied with security specifications alone \cite{hadar2018privacy}. Specifically, 41 participants (58.6\%) indicated that RDPRs require a combination of legal, technical (security), and procedural measures. Therefore, satisfying security requirements alone does not guarantee compliance with all RDPRs. \textit{`So privacy by design is considered at the very end and mostly security obligations are understood, not privacy ones whatsoever, not data protection, nor privacy, which are conflated with security''} (P 33).
 
 \begin{mdframed}
    Finding \findingsnumber: All regulations worldwide mandate security requirements for protecting personal data. However, security specifications alone are not sufficient to ensure compliance with these regulations.
\end{mdframed}

Most participants, when prompted about their reaction to a hypothetical data breach scenario, would consistently provide the same answer, as shown in \Cref{tab:challenges2}: They would check for the \textit{`Risk of data'} that was affected (30 participants; 42.9\%), which included identifying the type of \textit{`Personal data'} (18 participants; 25.7\%) affected, and categorize the \textit{`Type of incident'} (36 participants; 51.4\%). Then, they would follow the security policies, apply \textit{`Incident response'} measures (43 participants; 61.4\%), and fulfill \textit{`Notification'} requirements (50 participants; 71.4\%) set by the law, including those of data protection \textit{`Authority'} (44 participants; 62.9\%) and the \textit{`Data subject'} (36 participants; 51.4\%). 
In countries with a reporting requirement, 15 participants (21.4\%) stated that they would report within the specified \textit{`Time'} required by law.

43 participants (61.4\%) reported involving their \textit{`Incident Response'} or cybersecurity teams when available. 27 participants (38.6\%) gave details, saying they would immediately aim to \textit{`Contain'} the breach. Containment and control were identified as top priorities before proceeding to subsequent steps. After containing the breach, participants indicated they would gather information about the incident, particularly the type of data affected. \textit{``Has it been given, you know, what is the nature of the breach? What's the nature of the loss of control? So, what's the data? How have we lost control?''} (P 11).  The risk associated with the data is pivotal in determining the appropriate response, including notification requirements for both authorities and data subjects. As one participant explained\textit{`And then the response team should conduct a risk assessment regarding what was leaked. If the risk assessment determines that, for example, there is an impact on privacy, an impact on sensitive personal data, or on special categories of data, the organization must follow the measures required by law in this case,for example, notification to the agency''} (P 57)$^\ddagger$.

From these responses it is possible to deduce that legal practitioners require a document that maps and details the processing of personal information, including the \textit{`Type of Personal Data'}, applicable security policies, and general processing details. This document should be easily accessible and readable, so that lawyers and incident response teams to consult it during emergencies. Some of the countries researched mandate a record of processing activities (\textit{`RPA'}), discussed later in this section, which requires identification of the type of personal data processed, among other elements. This document should satisfy the aforementioned requirements. 

\begin{mdframed}
    Finding \findingsnumber: Legal practitioners require the following elements to react promptly to a data breach: (1) an incident response policy with clear line of command and response plans; (2) a document mapping and detailing the personal data processing, including the of type of personal data and processing details, such as access policies; (3) notification procedures with identified time frames, including the corresponding authorities and data subjects; and (4) risk assessment frameworks so lawyers can determine next steps.   
\end{mdframed}

Our analysis indicates the presence of additional subcodes to a lesser extent. One that was commonly agreed upon is the impact of data breaches on the organization's reputation. Ten participants mentioned the code \textit{`Consequences (reputation)'}. However, for space limitation we do not provide a discussion in detail, given the low number of responses coded with this code.

\paragraph{Children's data} 

A majority of participants (66; 94.3\%) discussed \textit{`Children/minor data'} in some manner. The Office of the United Nations High Commissioner for Human Rights emphasizes the digital well-being of children, who have specific, more stringent requirements \cite{childrenDigitalUN}. Notably, 43 participants (61.4\%) reported that some form of provision exists for minors' personal data, regardless of whether it is explicitly specified in the regulation. Our data triangulation concluded that 12 regulations establish requirements for minors in different ways. Therefore there is a discrepancy between what our participants state and our triangulation efforts.

The discrepancy between the high response rate of participants and the regulations may be explained by several factors. There are \textit{`Other regulation'} (mentioned by 18 participants, 25.7\%) that establish requirements for this type of data subject. Alternatively, there are \textit{`Interpretation'}(s) regarding childrens' personal data(4 participants, 5.7\%). All the countries in the sample have ratified the United Nations Convention on the Rights of the Child (UNCRC), except for Azerbaijan and Saudi Arabia\footnote{All EU countries have signed it, too.} \cite{RatificationCRC}. Additionally, regional and international organizations have issued legal documents addressing children's safety in digital environments, including privacy expectations, such as those from the EU \cite{commissionGuidelinesArt28}. 

To illustrate the discussion provided here, compare these excerpts on minors' data protection requirements from participants across different regulatory frameworks and continents. All their contributions lead to the same conclusion: regardless of where the requirement comes from, identification of minors is important for RDRPs. 
\begin{itemize}
    \item \textit{``It always was a concern from the very beginning except that gradually more laws have popped up, more design codes have popped up in the UK, California now''} (P 53);
    \item \textit{``In Singapore, because in our law itself, there's no definition for child or minor [...] one year ago, the PDPC issued advisory guidelines on the protection of children.  That's the first time that there were standalone advisory guidelines governing the use and processing of children's data.''} (P 22);
    \item \textit{``Regarding the personal data on minors and teenagers, the bill [signed upcoming law] establishes specifics rules and classifies such data, but not entirely, as  sensitive personal data''} (P 57)$^\ddagger$;
    \item \textit{``Yeah, yeah, it does. There are specific safeguards for minors [regarding the GDPR]''} (P 39).
    
\end{itemize}

These verbatims show that most participants agreed that a different type of data protection regime was necessary for minors' data. 18 participants (25.7\%) equated the management of this type of personal data to that of \textit{`High risk/Vulnerable'} individuals. To be more specific, 8 of them (11.4\%) would treat it as \textit{`Special Categories'} of personal data, while 7 participants (10.0\%) would treat minors' data as personal data coming from \textit{`Vulnerable'} individuals. Consequently, minors' personal data requires a specialized management, which usually has stricter requirements.

On another note, 7 participants (10\%) also expressed uncertainty (code \textit{`Not sure'}) about whether the data protection law included any provisions for minors. Finally, 5 participants also developed the idea of \textit{`Progressive Autonomy'}, in the sense that minors younger than 13 years are different from those between 13 and 18 years old. Therefore, their privacy expectations progress as they approach adulthood.

\begin{mdframed}
    Finding \findingsnumber: Most of legal practitioners consider children's data to be a higher-risk type of personal data. As such, it has different, more stringent requirements. The data protection requirements for children's data stem not only from data protection regulations but also from other legal instruments. Organizations must identify whether the data subject is a child.
\end{mdframed}

\paragraph{Documentation} 

Certain documents are essential for the processing of personal data. Privacy policies and privacy notices, though sometimes referred to by different terms, are referenced in all 21 regulations. A privacy policy is defined as an internal document that governs privacy and data protection practices. It is both an internal document of the organization and can also be provided to the general public. Privacy notices aim to inform users and data subjects about an organization's data processing activities before the organization captures and processes the personal data. Privacy notices are particularly significant when consent is collected, as they are required to inform the data subject. Regarding the content of privacy notices, as shown in \Cref{tab:summary-regulations-short}, the analysis of the different jurisdictions shows that the following information seems to be consistently requested (albeit with varying terminology):

\begin{itemize}
    \item Information identifying the controller and how to contact them (regardless of whether the differentiation between controller and processor existed);
    \item The specific purpose for processing the personal data and the lawful basis;
    \item Recipients or categories of recipients of the personal data;     
    \item Information about data subjects' rights and how to enforce them.
\end{itemize}

Other important documents are the Record of Processing Activities (RPA) and the Data Protection Impact Assessment (DPIA), which are mandated by 12 and 13 jurisdictions, respectively. These numbers are close enough to our minimum threshold of 14 regulations to be conceptualized as common, particularly for the DPIA. Given the importance of these documents, especially within the context of the SDLC, and the difficulty of writing them as an afterthought, we categorize them as common to avoid further work once the IS has been developed. Additionally, they are considered good practice in multiple regulations, and recommended by data protection experts. Across the regulations that mandate RPAS, this document aims to map all processed personal data, including the purpose, legal basis, and retention period, among other key information. The DPIA serves a different purpose: it must assess the risks of the data processing activities and evaluate the balance between the activity and the data subject's rights and freedoms. While specific requirements may vary across regulations, the overarching objectives remain consistent.

Registration with authorities for the processing of personal data is not commonly required across regulations.

\begin{mdframed}
    Finding \findingsnumber: Privacy policies are required in all jurisdictions. Privacy notices are also common, with a set of similar information across regulations. Data protection impact assessments are also required, even if the requirements differ.
\end{mdframed}

\subsection {SQ2: Which data protection requirements diverge across regulations, and how are they conceptually different?}

\paragraph{Controller versus processor}  
As discussed in the previous research questions, regulations identify similar actors but also differ. The biggest differences our participants highlighted were in nomenclature, which was difficult to translate into English, and in some of their duties. For example, one participant explained: \textit{``I'm not confident I have to translate this term handler because it's not an uncomfortable term to me ... So I will, in this case, I will just directly translate [...]  So Korean controllers are similar to EU controllers because they decide the processing purposes and means [... ] But the Korean personal data handlers differ from EU processors. So, handlers are employees under a controller's supervision''} (P 27).  

The most notable difference our participants highlight is the duties and responsibilities of processors. Controllers seemed mostly similar, but the duties of a processor could vary. \textit{``The other thing I would say is they also have a data processor, which is kind of like the data processor of GDPR, but a much smaller set of, I mean, I think they really reduced the number of obligations on the processor''} (P 15). Even how the relationship controller-processor is established differs. Under the GDPR, a contract must be in place; however, it is not the same every time. \textit{``But our code does not mandate a particular form for the controller or processor relationship between them, but in practice, controllers use a written agreement to give instructions and allocate responsibilities, but this is a compliance tool rather than a statutory requirement''} (P 30). 

In this sense, one participant made it clear that although specific aspects may differ between regulations, the general idea is similar. \textit{``I don't think the language, it doesn't, I don't think it copies the GDPR determining the purposes and means, but it's similarly, it conveys a similar idea''} (P 51).

\begin{mdframed}
    Finding \findingsnumber: Some duties from processors and controllers may differ between regulations. The establishment of the controller-processor relationship can also differ.
\end{mdframed}

\paragraph{Time frames} 

There is no uniform time frame for data subject rights, breach notifications, or similar events. For example, the GDPR in Art. 12(3) requires organizations to respond to a data subject rights request within 30 days, with a possible 30-day extension \cite{GDPR}. Azerbaijan's personal data protection law in Art. 12(4) requires 7 days, with an extension up to 15 days \cite{AzerbaijanDPLaw}. Brazil's LGPD in Art. 19(2) sets 15 days \cite{LGPD} and Japan's APPI in Art. 33(2) requires action without undue delay \cite{APPI}. Time frames for DSR vary from undue delay (such as Japan \cite{APPI}), reasonable time (Canada's PIPEDA, Schedule 1, Principle 9 \cite{PIPEDA}), 72 hours (Indonesia's PDP \cite{PDPLaw}), to 30 days with a possible 30-day extension (like the GDPR \cite{GDPR}, or the upcoming Chilean law \cite{Ley21719}). Therefore no common regulatory time frame exists. We provide a detailed reporting of this topic in our external repository.

Regarding data breach \textit{`Notification'}, although most participants (50; 71.4\%) suggested that one of the first actions they would take is to notify different actors of the breach, the time frames themselves also differ. As one participant said, \textit{``Following the rules of most data protection law, if the breach is up to a certain magnitude, then we must immediately inform the data protection authorities''} (P 1). The shortest time frame we found was  `as soon as possible and in the fastest medium' per Art. 14 (sexies) of the Chilean law \cite{Ley21719}. Other time frames include `as reasonable as possible on discovery' (Section 22(2), POPIA, South Africa \cite{POPIA}) or 72 hours (Art. 33 GDPR \cite{GDPR}) among other. Some data protection laws, like in Argentina \cite{Ley25326}, do not specify a time frame or process for breach notifications. We found that, in some regulations, the absence of specific time frame for data breach notification is set out in other regulatory documents, and not in the data protection regulations. Therefore, organizations must verify this requirement beyond data protection regulations

\begin{mdframed}
    Finding \findingsnumber: Most regulations set time frames for both data subject rights and data breach notification. However, time limits differ widely and may not be set in the data protection regulation.
\end{mdframed}

\paragraph{Data subject rights} 

We identify data subject rights that are not common across regulations: the right \textit{`To Object'}, \textit{`Data Portability'}, \textit{`Not to be subject to Automated decision-making'} and the right to \textit{`Erasure}, conceptualized as the \textit{right to be forgotten}.

\textit{`Data Portability'} was mentioned by 17 (24.3\%) participants. When analyzing the regulations, we found it in 8 regulations, with 1 regulation (China \cite{PIPL}) coded as partially as it did not follow our full conceptualization, in line with previous research indicating that this right is not widespread as it is new \cite{lim2025navigating}. This right requires that personal data be provided to data subjects in an interoperable manner across technologies, which presents multiple technical challenges  \cite{pandit2018gdpr,de2018right}, and it not widely adopted \cite{casalini2021}. As one participant puts it, \textit{``But like the right for portability definitely is a big question mark, especially because even the EU, no one knows what to do about it''}  (P 64).

\begin{mdframed}
    Finding \findingsnumber: The right to portability is new, not widely adopted across regulations yet, and presents technical challenges.
\end{mdframed}

The right to \textit{`Erasure'}, or the \textit{right to be forgotten}, is complex and nuanced, as it is both common across regulations and subject to variation. Generally, data that lacks a processing purpose or has been processed unlawfully should be deleted, as previously discussed. This conceptualization is referred to as the right to \textit{`Erasure'} in this research. However, this right includes multiple exemptions, and therefore, legal experts should verify them when dealing with this requirement.  

On the other hand, the idea of the \textit{right to be forgotten} emerged with the new conceptualization of this right under the GDPR \cite{BUNN2015336}. In particular, this right gained traction after the Google Spain SL, Google Inc. v Agencia Española de Protección de Datos, Mario Costeja González (2014) ruling \cite{G.vs.Spain}, which led to the right being enshrined in the GDPR \cite{BUNN2015336}. It not only mandates that the data be deleted but also that, if such data has been made public, it should not be visible once the right has been exercised. Multiple interpretations have been suggested over this right \cite{ambrose2013right,graux2012right,BUNN2015336}, including the distinction of `\textit{droit à l'oubli}', making it equal to erasure or even deleting data linked to oneself but not made accessible by oneself \cite{BUNN2015336}. We recognize the objective of this paper is not to discuss the conceptualization of this right, but we approach the \textit{right to be forgotten} as a more modern and distinct conceptualization of the right to erasure.  

This interpretation highlights that the \textit{right to be forgotten} varies across countries, as noted by several participants. In Latin American countries, historical experiences such as dictatorships have influenced a conceptualization of this right that differs from the GDPR. For instance, one participant stated, \textit{``We don't have that; we don't have the right to be forgotten, but the Constitutional Court, through its case law, has indeed recognized that right, and it says—this is just a small part of it—the right to data erasure'')}(P 67)$^\ddagger$. The Chilean regulation, for example, addresses the relationship between the right to be forgotten is not the same as erasure (or cancellation, per the authors in \cite{ortiz2021repensando}) while also being a European-centric right `transplanted' in the Chilean context \cite{ortiz2025derecho}. 

A participant from a Latin American country provided a detailed explanation: 

\textit{``For example, Article 17 on the right to be forgotten, that’s a load of nonsense that makes no sense at all in Latin America. And it’s not as if Latin America doesn’t recognize the right to have information deleted. The right to cancellation (erasure) exists, but it’s very different from the right to be forgotten [GDPR conceptualization], and it’s very different from putting a company to judge and  determine whether something should remain online or not, especially if you weren’t the one who created it in the first place. This issue arises with search engines indexing information that already exists on the Internet—which was the subject of a case against Google at one point\footnote{Referring to  \cite{G.vs.Spain}.} — but in Latin America, our reality is: `Wow! Censorship!' ''}(P 49)$^\ddagger$.

Similarly, other participants from North America also mentioned this difference: \textit{``But he didn't, the commissioner didn't want to go down that route because he sees the right to be forgotten as being more nuanced and balancing of free speech and all these things that sort of the framework that they now use in Europe''} (P 46). Alternatively, from the same country, \textit{``Very similar [referring to DSRs], but not the right to be forgotten. We have just in Quebec''}(P 48).  Some participants linked it to freedom of speech, \textit{``And we don't have a right to erasure, a right to be forgotten, because of that likely conflicting with the First Amendment''}  (P 51).

Participants from Asia expressed similar opinions: \textit{``There's no right to be forgotten under the Japanese law''} (P 20), or \textit{``But since GDPR brings new definitions, we don't have them, all of them. For example... for example, the right to be forgotten, is it? No, no''} (P 30). This finding is in line with what \citet{casalini2021} reports. As such, all regulations foresee mechanism for erasing data, but not in the conceptualization of the \textit{right to be forgotten}.

\begin{mdframed}
    Finding \findingsnumber: Although all regulations (and 31 participants mentioned it) have the right to erasure, its conceptualizations as \textit{the right to be forgotten} and its requirements differ, as the participants and specialized literature points out. Institutional and historical political culture of the jurisdiction plays a role in its conceptualization. The `right to be forgotten' remains mostly a European conceptualization according to our participants.
\end{mdframed}

Finally, the full conceptualization of the right \textit{`To Object'} is also not widespread across regulations. We do make a distinction between the right \textit{`To Revoke Consent'} and the right to \textit{`Object direct marketing'} and \textit{`To Object'} (general) to understand how different countries perceive this conceptualization. For example, for countries that impose requirements on \textit{`Direct marketing'} but do not address other aspects of the right to object, we classify them as partially for the right to \textit{`Object'}, as this right is only allowed in a very specific circumstance. In this sense, \Cref{tab:summary-regulations-short} shows that 10 countries provide the right to \textit{`Object'}, with 4 providing elements for direct marketing solely, and 12 do not foresee this right in any manner. Similarly, in the interviews, this right was in the lower bound of rights discussed, with 14 participants (20\%) discussing it.

\begin{mdframed}
    Finding \findingsnumber: Most personal data protection regulations do not generally include the right to object in their frameworks, apart from withdrawing consent. 
\end{mdframed}

Finally, most countries do not require data subject rights to be \textit{`Free of charge'}. In fact, only 8 require the right to access to be free, as already discussed in the previous question. For all the rights to be free, this number diminishes to 6 regulations, as reported by the code \textit{`All DSR free of charge'}. This finding can lead to future research to discuss if DSR should be free or not.

\begin{mdframed}
    Finding \findingsnumber: Most personal data protection regulations do not require that data subjects' rights be free of charge to exercise. 
\end{mdframed}

\paragraph{Legal basis - consent and derogations}

There are multiple legal bases across jurisdictions that are not common. Given the different possible legal bases, we do not list them all. In the document in our public repository we note when there are other legal bases. There are two particular elements to highlight, considering the GDPR as the referencing framework: \textit{`Legitimate interest'}, and whether \textit{`Consent'} should be \textit{`As the norm'} and hence the other legal bases as derogations. 

Over the first topic, \textit{`Legitimate interest'} does not appear to be a common legal basis. Although it appears in 12 regulations (as shown in \Cref{tab:summary-regulations-short}), it is not widespread. Indeed, according to one of our participants, \textit{``There is no such thing as a legitimate interest''} (P 54)$^\ddagger$.

Regarding whether \textit{`Consent'} should be \textit{`As the norm'} legal basis and the others possible legal basis derogations, it was noted in 7 regulations, which also participants from those countries highlighted.  This requirement implies that organizations within the scope of these regulations should opt for consent as the legal basis. If they prefer --- and are allowed --- to use another legal basis, organizations must present a reason why they are not using consent, as iti s a derogation. As two participants explained: \textit{``So most of the times, the legal basis will be consent. It's mandatory. Not mandatory, but it's like 90\% of the cases''} (P 52) and \textit{``The current system is based solely on consent, that's all there is. The rest is just a system of derogations, so it's extremely difficult to obtain personal data here. I mean, if you don't have the data subject's consent, that's it, good-bye''}  (P 67)$^\ddagger$.

This approach differs from that of the GDPR, where all legal bases are, to say in a manner, at the same level. In other words, organizations do not have to opt for \textit{`Consent'} as a legal basis and may choose what suits better for their processing goals.
\begin{mdframed}
    Finding \findingsnumber: Each jurisdiction may have its own legal basis that may not appear in other jurisdictions. Although it is increasingly appearing in regulations, a legitimate interest is still not widespread across jurisdictions. The choice of consent or other legal basis will depend on whether the conceptualization is done as an exception or has the same level as other legal bases. 
\end{mdframed}

\paragraph{Power Asymmetry and Vulnerability} 

52 (74.3\%) discussed this topic in some manner and most of them expressed uncertainty over its inclusion in the data protection regulation, frequently noting that only vague provisions, preambles, or interpretations from other instruments exist. Vulnerability in particular was discussed by 24 (34.3\%) participants, with most agreeing it is not a clear requirement. For instance, one participant stated: \textit{``It's interesting because the legislation, yes, it does mention, it does mention, but it doesn't provide specific sessions or provisions for that. But it brings this idea that the elderly should also be particularly considered''} (P 64). In some cases, vulnerability was referenced without a clear definition: \textit{``All I can say is that I think the Data Protection Act of Tanzania talks about, it mentions something about vulnerability, but it doesn't really define what vulnerability is''} (P 1).

Building on these perspectives, 12 participants (17.1\%) suggested that vulnerability could be considered a special category of data due to the associated risks in processing such information. For example, one participant noted: \textit{``I think you can infer them from the list of data that is classified as sensitive personal data''} (P 55). Other participants explicitly stated that vulnerability is included within special categories of data: \textit{``Yes, it mentions vulnerable individuals, but the only provision it actually make for so-called vulnerable or special information is under special personal information''} (P 5).

Triangulation with the main regulations and guidance revealed that it was not possible to code regulations on this subject consistently and in an academically rigorous manner. The terminology used across different regulations varied, and some terms risked being lost in translation. For example, one participant referenced: \textit{``So here it was mentioned that the process article 26, paragraph one said that the personal data protection, personal data processing of disability people is going to be carried out''} (P 19). This participant reveals that disability in their cultural context is conceptualized as a vulnerability. Although all relevant regulations were reviewed and some references to vulnerability were identified, it remains possible that important aspects were overlooked. Given that even national data protection experts expressed uncertainty (as indicated by 10 participants, under \textit{`Not sure'}) and most relied on interpretation, there was insufficient information for us to code the regulatory framework. Future research could address this gap through comparative regulatory analysis.

Shifting focus to the broader legal context, as \citet{malgieri2020concept} has argued, vulnerability is part of the GDPR and a data protection requirement. The power asymmetries between the actors entail specific requirements that would not exist in symmetrically structured power structures, such as employer-employee relationships. One participant provided an extensive analysis of what vulnerability means in the GDPR context, which summarizes our discussion:  

\textit{``The short answer is the GDPR just mentions vulnerability in a recital. I think it's 72 about high risk [...]  There's not a label of vulnerability that you wake up as vulnerable, you go to eat as vulnerable, you have sex as vulnerable, you go to bed as vulnerable, but it's more a contextual layered notion. So in certain data processing relationships, there might be a situation of higher risks for your freedom, autonomy, dignity, and fundamental rights. And this turns into you being considered vulnerable in that situation''} (P 39).

In summary, vulnerability, although not yet a common requirement across all data protection regimes, appears to be attracting attention.

\begin{mdframed}
    Finding \findingsnumber: Vulnerability is not commonly included as a data protection requirement, but it seems to be gaining traction. 
\end{mdframed}

\paragraph{International Transfers}

Finally, all the regulations we analyzed allow international transfers of personal data, in some form. Of the participants interviewed, 51 (72.9\%) described how these transfers could occur, which matches our analysis that all 21 jurisdictions allow cross-border transfers. There is little commonality on how these transfers work. From the qualitative analysis, the most common code is \textit{`Similar to the GDPR'}. However, when we triangulate with the literature and the regulation, there seem to be differences. For example, one participant notes: \textit{``Our regulations regarding countries that our country considers suitable, oh surprise, are practically the same as Europe’s. And, in fact, we added England at the same time as Europe; well, n... of course, Europe had added it earlier, and we did so later, so, well, we’re keeping up and following their footsteps''} (P 68)$^\ddagger$. 

Most regulations differ in when and how transborder flows can occur, imposing varying requirements. Notably, regulations with strong Brussels Effects follow guidance akin to the GDPR, as evidenced by the code \textit{`Similar to the GDPR'} being discussed by 26 (37.1\%) participants. In these cases, possible mechanisms to transfer data outside a jurisdiction include adequacy decisions or lists of safe third countries, SCC, BCR, treaties, certifications, and consent as an exception. Furthermore, some participants went beyond and recognized that their jurisdictions would follow the adequacy decision provided by the EU Commission, as already discussed. 

Building on this divergence, jurisdictions differ significantly in specific processes, as evidenced by our regulatory analysis, in which almost none of our codes reach 50\% or higher, except for \textit{`BCR'}, that has been in coded in 11 regulations. This situation is also recognized by the OECD \cite{casalini2021}, which raises concerns about its economic impact and calls for stronger cooperation. Although some countries, such as Colombia and Brazil, appear similar to the GDPR, they differ in that they allow consent as a basis for transferring personal data outside their jurisdictions, whereas the GDPR treats it as a derogation. One participant described, \textit{``Consent, which, unlike the GDPR, is not an exceptional hypothesis, so you could use consent regularly, it's not something residual''} (P 65). As another participant from Canada explains:\textit{``For sure, when you send the data outside of Canada. But at the same time, it's not mandatory to have a data transfer agreement under GDPR''} (P 45).

\begin{mdframed}
    Finding \findingsnumber: TPDFs are usually allowed within jurisdictions, as long as they fulfill the requirements set by the law. However, these requirements and the legal basis for transferring data outside differ. 
\end{mdframed}

\section{Data Protection Stories} \label{sec:dpo-stories} 

Based on these findings, we operationalize them into requirements. A widespread artifact to operationalize requirements within agile frameworks for the SDLC is user stories, as discussed in \Cref{sec:related-work}. We created 74 user stories based on the most important elements identified here, which we will refer to as DPO stories. We omitted security requirements from these DPO stories because there is ample prior work and literature on security RE that can be used in combination with our contribution \cite{seeba2025evaluating,gharib2024us4usec,bartolini2019gdpr}. These stories are intended primarily for use by DPOs within organizations but can also be used by any other stakeholders who need to apply RDPRs. The complete list of DPO stories is available in our public repository and in \Cref{subsec:user-stories-methods}. 

The DPO stories are divided into two categories: common and uncommon. Additionally, they are organized by theme of analysis rather than by epic user stories. Common requirements include: personal data, actors, legal bases, minors’ personal data, and documentation. The uncommon requirements are divided as follows: time frames, certain legal bases, vulnerability, and how to do TPDFs. The actors in the DPO stories are based on our conceptualization of `actors' in \Cref{sec:analysis}; that is to say, data subjects, data protection officers (DPO), controllers and processors. Since these DPO stories are intended primarily for private organizations, we do not provide any for data protection authority(ies). Similarly, when a requirement is important for both the controller and the processor, we use the type of actor organization; meaning both the controller and the processor must satisfy the goal. 

Due to space limitations, our analysis focuses on the common requirements here, but the DPO stories for the uncommon requirements are available in the \Cref{annex:user-stories}. The detailed process for the constructing the DPO is explained in \Cref{subsec:user-stories-methods}. 

We provide three examples of different DPO stories, each part with a different theme. Furthermore, as part of our contribution, we provide a preliminary classification for each DPO story within the EA and SDLC frameworks in brackets (e.g., [EA, SDLC]). This preliminary classification can give DPOs or whoever is using the artifact a hint about where to place emphasis on this user story. This is a preliminary classification that needs to be validated by domain experts and can be modified by the end-user as they see fit, therefore serving as a suggestion.

\begin{mdframed}
    \begin{itemize}
        \item Actors: As a controller, I want to verify that personal data has been deleted or returned by the processor when a contract ends, so that the personal data cannot be reused. [Business, Operation]
        \item As a data subject, I want to withdraw my consent for the processing of my personal data by the organization in a simple and accessible manner. [Information, Design/Operation]
        \item As the DPO, I want to write the privacy notice in plain language, so that data subjects can understand it. [Application, Requirement]
    \end{itemize}
\end{mdframed}

When classifying the DPO stories, the two most prominent areas within the EA framework are both business and information. As a result, RDPRs aim to set requirements for the governance of personal data, rather than the design of an IS. For example, DPO stories on `Legal Bases’ primarily concern business because the legal basis for processing personal data depends on the purpose of such processing, which aligns with an organization's business objectives. Similarly, the inclusion of other actors (or stakeholders) within the IS arises from business necessity rather than from an application or technology. 

Regarding information, the DPO stories of DSR, minors, and personal data are mostly classified within this section. For example, DSR, as they are linked to the manipulation of personal data, most relate to the modification (or general processing) of such data. Although there are DPO stories related to the application level (such as how the privacy notice should be presented), most were at the information level. 

The only theme that does not follow this trend is documentation, which has DPO stories at every level of the EA, and the SDLC. 

Moving into the SDLC, all DPO stories are requirements; that is their purpose. However, some of these requirements play a more important role in different areas of the SDLC, such as goal setting, design, and operations. As such, we provide a preliminary classification when a DPO story has increased importance within the SDLC. The only SDLC phase that does not have any DPO story is the deployment phase. Then, on the lower bound testing has 2 DPO stories, there are 3 in architecture, and 6 in testing. This demonstrates how the RDPRs, once again, give organizations the freedom to design their IS as long as they respect the RDPRs\footnote{This does not include security requirements.}.

Only two themes within our DPO stories showed a tendency towards a particular phase of the SDLC: the majority of DPO stories for DSR are related to the operational phase, and minor data are all requirements. DSR is highly linked to the operational aspect of the SDLC, as they are triggered once the software has already been deployed and is processing personal data from data subjects. Minor data is concentrated at the requirements level of the SDLC, as it is important to know whether such data will be processed before development begins, since this will affect the design, architecture, and other aspects of the IS. 

The DPO stories can be used by privacy experts or any other person involved in the IS development. Their idea is to help guide teams on data protection aspects, regardless of where they are based. That is to say, these DPO stories should enable a minimum level of compliance across regulations. The classification within the EA and SDLC we present here is a preliminary suggestion that experts can modify as they see fit. 

Based on the results previously presented, we propose a set of good practices based on the catalog of data protection user stories. The whole set of the so-called `DPO stories’ we have provided is available in our public repository  the organization to use. 

 \section{Threats to validity, limitations and future work} \label{sec:threats}
This research does not aim to be an exhaustive comparison of regulations worldwide, nor to exemplify all regulatory data protection requirements that may exist, which may span different laws. There are different limitations to this work, which we discuss in detail.

This paper should be understood within its objective: to identify common conceptualizations of RDPRs worldwide, based on interviews and legal research, and to provide user stories (which we call DPO stories here) that organizations can use to conceptualize the satisfaction of the requirements. In this sense, and in light of \cite{lewis2003generalising}, we consider that our article demonstrates strong theory-building regarding commonalities in data protection requirements, particularly considering the Brussels effect theory in this field.

In qualitative research, reliability and validity may differ from those in the natural sciences \cite{lewis2003generalising,lincoln}. We also use other terms that may better fit our work: credibility, transferability, dependability, trustworthiness, and confirmability \cite{ahmed2024pillars,lincoln}. We justify our limitations by asking whether there is sufficient evidence to assess the confidence in our results, their applicability, repeatability, and possible biases \cite{anney2014ensuring}, which are crucial to qualitative research methods.

To help with trustworthiness and credibility alongside reliability, we provide a thick description of the method \cite{anney2014ensuring} along with details and recounts of the process \cite{lewis2003generalising,ahmed2024pillars}. We provide as much data as possible while respecting participant anonymization preferences. Some interview transcripts are openly available, while others can be accessed upon request. We also share supporting materials, including the invitation email, codebook, interview questions, and automated analyses. However, some materials cannot be disclosed to protect participant identities. 

To enhance credibility, trustworthiness, and reliability, we employed additional strategies with participants. Prior to interviews, we exchanged emails, facilitated initial communication to establish rapport, and engaged in rapport-building interactions \cite{ahmed2024pillars,anney2014ensuring,babbie2020practice}. In some cases, rapport had to be built quickly due to participants' limited availability \cite{babbie2020practice}, which was suboptimal; with others, rapport developed gradually through ongoing pre- and post-interview conversations, reflecting prolonged engagement \cite{anney2014ensuring,ahmed2024pillars}. Such engagement indicates participants were comfortable and communicated openly.

To minimize potential biases that could affect reliability and credibility, we took several measures during coding. We started with codes based on the GDPR, as the Brussels effect theory suggests that most of these requirements are widely adopted, per our method \cite{fife2024deductive}. We also used previous research for codes \cite{hadar2018privacy,sutcliffe2010collaborative}. To further reduce bias, we validated the codes with lawyers from various countries, who provided feedback and suggested modifications, engaging participants \cite{westbrook1994qualitative}. This also aimed to increase the face validity of our codes.

Per DQA, an early analysis of interviews allowed the first author to immerse themselves in the raw data and identify codes that may have been biased and needed refactoring \cite{fife2024deductive}. All codes were discussed with the other authors, with special attention to those that the lead author found may be biased, and were modified accordingly. Inductive codes were also incorporated when necessary, as permitted by DQA, \cite{fife2024deductive}, so we did not lose relevant data in our analysis. Although this strategy is not fail-proof, the constant evolution and discussion of the codes, along with external validation by domain experts, help us reduce bias. Nevertheless, the codes may contain certain biases, considering that it took the Brussels effect in data protection as a starting point per our methodology \cite{fife2024deductive}. Future research could develop the codebook using another theory and compare the results to determine whether they are similar. The codebook is publicly  at \url{https://doi.org/10.5281/zenodo.15166882}.  

To further mitigate coder bias, at least two authors independently coded the data, reviewed and re-coded each other's work, discussed, and agreed on the coding, similar to \cite{larusdottir2024ucd}. This approach resembles a stepwise replication process, which contributes to dependability  \cite{anney2014ensuring}. Consistent, open communication was maintained throughout the coding process to resolve discrepancies by consensus.

Finally, we shared our results with 7 participants to perform a member check \cite{westbrook1994qualitative,anney2014ensuring}. We did not receive major comments on our analysis. Rather, the participants commented on the style of writing and word choices, which were duly noted and documented by the lead researcher. This process further addresses potential researcher bias.

The lead author maintained an audit trail throughout the study, including interview notes, anonymized transcripts, memos, and records of author meetings. Audio recordings were destroyed after transcription. While this documentation is not publicly available for confidentiality reasons, it may be requested and supports transparency and reliability assessments.

Once we have coded the interviews and obtained preliminary results to increase reliability and validity, and to provide greater precision and robustness to our conclusions, we triangulate our results \cite{westbrook1994qualitative,lewis2003generalising}. In particular, we use different approaches to triangulation: method, data, and investigator triangulation. The latter has already been described, as multiple researchers analyzed the data \cite{anney2014ensuring}. The methods and data triangulation span from our regulatory analysis \cite{lewis2003generalising,anney2014ensuring}. By having another source of information, analysed through another method, we are able to provide more robust and credible findings \cite{westbrook1994qualitative}. Future work could also conduct quantitative analysis using surveys to provide further evidence and more precise results.

Regarding the study's use of purposive sampling, this decision could also limit the reliability and validity of our conclusions from a natural science perspective \cite{lewis2003generalising}. From a qualitative perspective, we sought rich insights from lawyers as domain experts. By engaging with experts from multiple countries, we believe our findings can be transferable across contexts involving data protection requirement \cite{anney2014ensuring,ahmed2024pillars,lewis2003generalising}. 

We aimed to have as much diversity within the data protection realm as possible: academics, practitioners, policy makers, different levels of expertise, among others \cite{lewis2003generalising}. Although we do not report this data granularly due to the risks of re-identification, we try to provide as much detail as safely possible.
An empirical evaluation of our findings and DPO stories in real-world settings is beyond this study’s scope. Future research could assess their practical usefulness through technical action research.

As a qualitative study, our analysis does not include statistical validation. Future work could use probabilistic sampling and factor analysis to further test and validate our findings.

Data protection regulations evolve over time, and our findings reflect a snapshot of requirements between 2025 and early 2026. While this may limits their long-term validity, the high-level nature of the topics can suggest that many conclusions may remain relevant.
This research was conducted over 15 months, during which regulatory and political changes may have influenced the results. For example, some participants in Chile referred to previous legislation rather than newly enacted rules.
To mitigate this historical validity concern, we triangulated interview data with regulatory coding and identified regulations undergoing significant changes. Nevertheless, some regulations may have changed between data collection and publication.

\section{Conclusion} \label{sec:conclusion}

Regulatory data protection requirements have become widespread worldwide, and organizations must comply with them. In this paper, we identified which requirements across regulations shared a common conceptualization and which differed. These results were obtained through a deductive qualitative analysis of 70 interviews we conducted with data protection experts from various countries. We triangulated these results with the analysis of 21 regulations, using systematic content analysis for legal artifacts.

Our results identify a list of requirements that have common conceptualizations, such as the definition of personal data, data subject, controller, which types of actors are defined in the regulations, some data subject rights, security requirements, documentation needed, among others. Similarly, there is a set of requirements with different conceptualizations, such as time frames, power asymmetries, or legal bases such as legitimate bases. As such, although the regulations may differ over time, certain aspects follow similar reasoning and logic, which we identify and present as conceptualizations. It is precisely this distinction, between what converges and what diverges across jurisdictions, that organizations must resolve to manage transborder personal data flows: shared conceptualizations can be addressed through common controls, while divergent ones require jurisdiction-specific treatment. Our work provides organizations with this map of convergence and divergence as a foundation for managing TPDF compliance.

With the identified conceptualizations, we provide a list of user stories, which we call DPO stories, that operationalize these requirements into actionable information-system actions. Organizations can use this artifact to discuss regulatory data protection requirements in a jurisdiction-agnostic way. The DPO stories are divided between the requirements we found to be common and those that differ.

To finish by quoting one of our participants, although data protection regulations may differ, the logic behind them is similar, which is what we have tried to convey in this research through our conceptualization and user stories. 

\textit{``The problem is that data protection law is not universal. There may be regional ones; there are some global standards, but there are also national legislations. And therefore, there are differences in between, and you know better than me because you do what you do in your research. But there is some logic, which is that even if the specifications may be different between different jurisdictions, different countries, the logic does not change''} (P 41). 

\section{Data availability statement} \label{sec:data-availability-statement}

Due to the qualitative nature of this study and respecting the consent and privacy of the participants, not all interviews can be shared. The data publicly accessible is available at \url{https://doi.org/10.5281/zenodo.15166882} (DOI: 10.5281/zenodo.15166882)
or upon request of the first author or corresponding authors, depending on the consent. In order to provide maximum transparency, we provide other materials in the repository, such as the codebook, interview questions, and other material deemed relevant. 

\section{Acknowledgment and funding}

This project has received funding from the European Union’s Horizon 2020 research and innovation programme under the Marie Skłodowska-Curie Actions grant agreement 101081455. The authors would like to thank Dr. Marius Lombard-Platet for his invaluable help and advice, and Prof. Oscar Pastor, Prof. Georg Mein and Prof. Gabriele Lenzini for their support and advice. 

\section{AI usage}

This article has used a paid version of Grammarly to correct and fix spelling mistakes. 
The codes definition in the codebook was written by the author, and they used Grammarly and Chat GPT 4o for identifying spelling and grammar mistakes. If consented by the participant, interviews were transcribed using AIKO, which was then proof-read and corrected to ensure accuracy of the transcription. The authors used Claude Opus to convert a list of URLs of various data regulations into a list of BibTeX entries for said URLs and manually checked and corrected the result. The authors have no other disclaimer on the usage of AI and all errors remain the authors'. 

 \bibliographystyle{elsarticle-num-names} 
 \bibliography{bibliography.bib}

\appendix

\section{Positionality statement} \label{app:positionality-statement}

The authors of this article have different academic backgrounds, from economics, computer science, engineering, informatics, law, psychology and political science. The leading (first) author has a decade of experience on security and privacy issues, working both in the private sector, academia, and activism, while also having a formal education in computer science and political science. Other authors have had private sector experience in privacy, cybersecurity, software development, and human-computer interaction. More than one author has contributed and worked with international organizations and governments on technology policy-building. 

\section{Final list of countries for interviews and regulatory analysis} \label{app:list-countries}
The countries are presented in alphabetical order.

Interview participant expertise: Argentina, Australia, Azerbaijan, Brazil, Canada, Chile, China, Colombia, Egypt, France (EU), Germany (EU), India, Indonesia, Italy (EU), Japan, Luxembourg (EU), Mexico, Saudi Arabia, Singapore, South Africa, South Korea, Tanzania, Turkey, United Kingdom, United States, and European Union (any country). 

Country regulations: Argentina, Australia, Azerbaijan, Brazil, Canada, Chile, China, Colombia, Egypt, European Union, India, Indonesia, , Japan,  Mexico, Saudi Arabia, Singapore, South Africa, South Korea, Tanzania, Turkey, and  United Kingdom.

\section{Detailed sample of participants}
\label{app:participants-details}

\begingroup  
\renewcommand{\arraystretch}{0.72} 

\footnotesize
\begin{longtable}{ ccc }
    \caption{Full list of participants of the interviews, per continent and their jurisdiction of expertise. Multiple legal experts hold knowledge and certifications regarding the GDPR. Junior = 2 - 3 years experience; Intermediate = 4 - 8 years experience; Expert = +8 years experience} \label{tab:part-long} \\

Participant & Experience & Continent of residence and expertise(s)        \\ \hline 
\endfirsthead

\multicolumn{3}{c}%
{{\tablename\ \thetable{} -- continued from previous page}} \\
Participant & Experience & Continent of residence and expertise(s)        \\ \hline 
\endhead

\hline \multicolumn{3}{r}{{Continued on next page}} \\
\endfoot

\endlastfoot

    Part. 1 & Expert  & Africa and Europe\\ 
    Part. 2 & Expert  & Africa and Europe\\ 
    Part. 3 & Intermediate  & Africa and Europe\\ 
    Part. 4 & Intermediate  & Africa and Europe\\ 
    Part. 5 & Expert  & Africa and Europe\\ 
    Part. 6 & Expert  & Africa and Europe\\ 
    Part. 7 & Expert  & Africa and Europe\\ 
    Part. 8 & Expert  & Africa \\ 
    Part. 9 & Intermediate  & Africa \\ 
    Part. 10 & Intermediate  & Asia and Oceania \\ 
    Part. 11 & Expert  & Asia and Oceania \\ 
    Part. 12 & Intermediate  & Asia and Oceania \\ 
    Part. 13 & Expert  & Asia and Oceania \\ 
    Part. 14 & Expert  & Asia and Oceania \\ 
    Part. 15 & Intermediate  & Asia and Oceania\\ 
    Part. 16 & Intermediate  & Asia and Oceania \\ 
    Part. 17 & Junior & Asia and Oceania \\ 
    Part. 18 & Expert & Asia and Oceania \\ 
    Part. 19 & Intermediate  & Asia and Oceania \\ 
    Part. 20 & Expert  & Asia and Oceania and \\ 
    Part. 21 & Expert  & Asia and Oceania and North America\\ 
    Part. 22 & Expert  & Asia and Oceania and North America \\ 
    Part. 23 & Expert  & Asia and Oceania and North America\\ 
    Part. 24 & Expert  & Asia and Oceania and Europe\\ 
    Part. 25 & Expert  & Asia and Oceania and Europe \\ 
    Part. 26 & Expert  & Asia and Oceania and Europe \\ 
    Part. 27 & Expert  & Asia and Oceania and Europe \\ 
    Part. 28 & Intermediate  & Asia and Oceania and Europe \\ 
    Part. 29 & Expert  & Asia and Oceania and Europe \\ 
    Part. 30 & Expert  & Asia and Oceania and Europe \\ 
    Part. 31 & Expert  & Europe and Asia and Oceania \\
    Part. 32 & Expert  & Europe \\ 
    Part. 33 & Expert & Europe \\
    Part. 34 & Expert & Europe \\
    Part. 35 & Expert & Europe \\
    Part. 36 & Expert & Europe \\    
    Part. 37 & Expert & Europe \\
    Part. 38 & Expert & Europe \\
    Part. 39 & Expert & Europe \\
    Part. 40 & Expert & Europe \\
    Part. 41 & Expert & Europe \\
    Part. 42 & Expert & Europe \\
    Part. 43 & Expert & Europe \\
    Part. 44 & Expert & Europe \\
    Part. 45 & Expert & North America and Europe \\
    Part. 46 & Expert & North America and Europe\\
    Part. 47 & Expert & North America and Europe\\
    Part. 48 & Expert & North America and Europe\\
    Part. 49 & Expert & North America and Europe\\
    Part. 50 & Expert & North America and Europe\\
    Part. 51 & Expert & North America and Asia and Oceania\\
    Part. 52 & Expert & North America and South America,\\
    Part. 53 & Expert & North America, South America, and Europe \\
    Part. 54 & Expert & South America \\
    Part. 55 & Expert & South America \\
    Part. 56 & Expert & South America \\
    Part. 57 & Expert & South America \\
    Part. 58 & Intermediate & South America \\
    Part. 59 & Expert & South America and Europe\\
    Part. 60 & Intermediate & South America and Europe \\
    Part. 61 & Expert & South America and Europe \\
    Part. 62 & Intermediate & South America and Europe \\
    Part. 63 & Expert & South America and Europe \\
    Part. 64 & Expert & South America and Europe \\
    Part. 65 & Expert & South America and Europe \\
    Part. 66 & Expert & South America and Europe \\
    Part. 67 & Expert & South America and Europe \\
    Part. 68 & Expert & South America and Europe\\
    Part. 69 & Expert & Africa and Europe \\ 
    Part. 70 & Expert & Europe \\
    \hline
\end{longtable}

\endgroup

\section{Detailed regulation notation} \label{annex:detailed-analysis-regulation}

\begin{landscape}
\centering
\scriptsize
\begin{longtable}{l|*{21}{@{}c@{}}}

\caption{Existence of the codes in personal data and data subject rights in the main data protection regulation or other important legal instruments. For a more detailed table, including regulations analyzed and articles, please go to the dataset.} \\
&\rot{Argentina} &\rot{Australia} &\rot{Azerbaijan}&\rot{Brasil}&\rot{Canada}&\rot{Chile}&\rot{China}&\rot{Colombia}&\rot{Egypt}&\rot{EU}&\rot{India}&\rot{Indonesia}&\rot{Japan}&\rot{Mexico}&\rot{Saudi Arabia} &\rot{Singapore}&\rot{South Africa}&\rot{South Korea} &\rot{Tanzania} &\rot{Turkey} &\rot{UK}\\
\hline\hline \endfirsthead

\multicolumn{22}{c}{{\tablename\ \thetable{} -- continued from previous page}} \\
&\rot{Argentina} &\rot{Australia} &\rot{Azerbaijan}&\rot{Brasil}&\rot{Canada}&\rot{Chile}&\rot{China}&\rot{Colombia}&\rot{Egypt}&\rot{EU}&\rot{India}&\rot{Indonesia}&\rot{Japan}&\rot{Mexico}&\rot{Saudi Arabia} &\rot{Singapore}&\rot{South Africa}&\rot{South Korea} &\rot{Tanzania} &\rot{Turkey} &\rot{UK}\\ \hline\hline
\endhead

\multicolumn{22}{c}{\cellcolor{grey!60}Conceptualizations of personal data} \\
{Legal persona} &  \CIRCLE & \Circle & \Circle & \Circle & \Circle & \Circle & \Circle & \Circle & \Circle & \Circle & \Circle & \Circle & \Circle & \Circle & \Circle & \CIRCLE\ & \Circle & \Circle  & \Circle & \Circle & \Circle\\

Natural persona & \CIRCLE & \CIRCLE& \CIRCLE& \CIRCLE& \CIRCLE& \CIRCLE& \CIRCLE& \CIRCLE& \CIRCLE& \CIRCLE& \CIRCLE& \CIRCLE& \CIRCLE& \CIRCLE& \CIRCLE & \CIRCLE & \CIRCLE & \CIRCLE & \CIRCLE & \CIRCLE & \CIRCLE \\

\hspace{1em} $\hookrightarrow$ Alive & \Circle & \CIRCLE & \CIRCLE & \CIRCLE & \Circle & \CIRCLE & \Circle & \Circle & \Circle & \CIRCLE &\Circle & \Circle &\CIRCLE & \Circle & \CIRCLE & \CIRCLE & \CIRCLE & \CIRCLE & \Circle & \CIRCLE & \CIRCLE \\

\hspace{1em} $\hookrightarrow$ Deceased  & \Circle & \Circle & \LEFTcircle  & \Circle &\LEFTcircle & \LEFTcircle & \LEFTcircle & \Circle & \Circle & \Circle & \LEFTcircle & \Circle &\Circle & \Circle & \LEFTcircle & \LEFTcircle & \Circle & \Circle   &\Circle & \Circle & \Circle \\

Identified & \CIRCLE  & \CIRCLE  & \CIRCLE  & \CIRCLE  & \CIRCLE  & \CIRCLE  & \CIRCLE  & \CIRCLE  & \CIRCLE  & \CIRCLE  & \CIRCLE  & \CIRCLE  & \CIRCLE  & \CIRCLE  & \CIRCLE  & \CIRCLE  & \CIRCLE  & \CIRCLE  & \CIRCLE & \CIRCLE & \CIRCLE \\

Identifiable & \CIRCLE & \CIRCLE & \CIRCLE & \CIRCLE & \CIRCLE & \CIRCLE & \CIRCLE & \CIRCLE & \CIRCLE & \CIRCLE & \CIRCLE & \CIRCLE & \CIRCLE & \CIRCLE & \CIRCLE & \CIRCLE & \CIRCLE & \CIRCLE & \CIRCLE &  \CIRCLE  & \CIRCLE \\

Special categories (SP) & \CIRCLE & \CIRCLE& \CIRCLE& \CIRCLE& \CIRCLE& \CIRCLE& \CIRCLE & \CIRCLE & \CIRCLE & \CIRCLE & \LEFTcircle & \CIRCLE & \CIRCLE & \CIRCLE & \CIRCLE & \LEFTcircle & \CIRCLE & \CIRCLE & \CIRCLE & \CIRCLE & \CIRCLE \\

\multicolumn{22}{c}{\cellcolor{grey!60}Individual rights} \\

To be informed & \CIRCLE &\CIRCLE &\CIRCLE &\CIRCLE &\CIRCLE &\CIRCLE &\CIRCLE & \CIRCLE & \CIRCLE & \CIRCLE & \CIRCLE & \CIRCLE & \CIRCLE & \CIRCLE & \CIRCLE & \CIRCLE & \CIRCLE & \CIRCLE & \CIRCLE & \CIRCLE & \CIRCLE\\

To Access  & \CIRCLE  & \CIRCLE  & \CIRCLE  & \CIRCLE  & \CIRCLE  & \CIRCLE  & \CIRCLE  & \CIRCLE  & \CIRCLE  & \CIRCLE  & \CIRCLE  & \CIRCLE  & \CIRCLE  & \CIRCLE  & \CIRCLE  & \CIRCLE  & \CIRCLE  & \CIRCLE  & \CIRCLE & \CIRCLE & \CIRCLE\\

\hspace{1em} $\hookrightarrow$ Free of fee\footnote{This does not include post services, if ask in physical format.} & \CIRCLE & \Circle & \CIRCLE & \CIRCLE & \LEFTcircle & \LEFTcircle & \LEFTcircle & \LEFTcircle & \Circle &  \CIRCLE & \Circle &  \CIRCLE & \Circle & \CIRCLE & \Circle & \Circle & \Circle & \CIRCLE & \Circle & \CIRCLE & \Circle  \\

To Rectification  & \CIRCLE & \CIRCLE & \CIRCLE & \CIRCLE & \CIRCLE & \CIRCLE & \CIRCLE & \CIRCLE & \CIRCLE & \CIRCLE & \CIRCLE & \CIRCLE & \CIRCLE & \CIRCLE & \CIRCLE & \CIRCLE & \CIRCLE & \CIRCLE & \CIRCLE  & \CIRCLE & \CIRCLE\\

To Erasure ($\neq$ forgotten) & \CIRCLE & \CIRCLE & \CIRCLE & \CIRCLE & \CIRCLE & \CIRCLE & \CIRCLE & \CIRCLE & \CIRCLE & \CIRCLE & \CIRCLE & \CIRCLE & \CIRCLE & \CIRCLE & \CIRCLE & \CIRCLE & \CIRCLE & \CIRCLE & \CIRCLE & \CIRCLE  & \CIRCLE \\

To Restrict Processing & \CIRCLE & \Circle & \CIRCLE  & \CIRCLE & \Circle & \CIRCLE  & \CIRCLE & \Circle & \CIRCLE & \CIRCLE &\Circle & \CIRCLE  & \CIRCLE & \Circle & \CIRCLE  & \Circle & \CIRCLE  & \CIRCLE  & \CIRCLE & \Circle & \CIRCLE  \\

To Data Portability & \Circle & \Circle & \Circle & \CIRCLE & \Circle & \CIRCLE  & \LEFTcircle & \Circle & \CIRCLE & \CIRCLE & \Circle & \CIRCLE & \Circle & \Circle & \CIRCLE  & \Circle & \Circle & \CIRCLE & \Circle & \Circle & \CIRCLE \\

Opt-out/Object & \Circle & \LEFTcircle & \CIRCLE & \CIRCLE & \Circle & \LEFTcircle & \CIRCLE & \Circle & \CIRCLE & \CIRCLE &\Circle & \CIRCLE & \LEFTcircle & \CIRCLE & \LEFTcircle &\Circle & \CIRCLE & \Circle & \CIRCLE & \Circle & \CIRCLE\\

\hspace{1em} $\hookrightarrow$ Object Direct Marketing & \Circle & \CIRCLE & \Circle & \Circle & \Circle & \CIRCLE & \CIRCLE & \Circle & \CIRCLE & \CIRCLE &\Circle & \Circle & \CIRCLE & \Circle & \CIRCLE &\Circle & \CIRCLE & \Circle & \CIRCLE & \Circle & \CIRCLE\\

To Revoke Consent  & \CIRCLE & \CIRCLE & \CIRCLE & \CIRCLE & \CIRCLE & \CIRCLE & \CIRCLE & \CIRCLE & \CIRCLE & \CIRCLE & \CIRCLE & \CIRCLE & \Circle & \CIRCLE & \CIRCLE & \CIRCLE & \CIRCLE & \CIRCLE & \CIRCLE & \CIRCLE & \CIRCLE\\

\begin{tabular}{@{}l@{}}Not to be Subject to\\ \,\, automated  Decision-Making\end{tabular} & \Circle & \Circle & \CIRCLE & \CIRCLE & \Circle & \CIRCLE & \CIRCLE & \Circle & \Circle & \CIRCLE & \Circle & \CIRCLE & \Circle  & \Circle  & \Circle & \Circle & \CIRCLE & \CIRCLE & \CIRCLE  & \CIRCLE & \CIRCLE \\

DS provides identity& \CIRCLE & \LEFTcircle & \CIRCLE & \Circle & \CIRCLE & \CIRCLE & \Circle & \CIRCLE & \Circle  & \CIRCLE  & \CIRCLE & \Circle  & \CIRCLE  & \CIRCLE  & \CIRCLE  & \CIRCLE  & \CIRCLE  & \CIRCLE  & \CIRCLE  & \CIRCLE & \CIRCLE\\

Complains  & \CIRCLE& \CIRCLE& \CIRCLE& \CIRCLE& \CIRCLE& \CIRCLE& \CIRCLE& \CIRCLE& \CIRCLE& \CIRCLE& \CIRCLE& \CIRCLE& \CIRCLE& \CIRCLE& \CIRCLE& \CIRCLE& \CIRCLE& \CIRCLE& \CIRCLE& \CIRCLE & \CIRCLE\\

Time Frame  & \CIRCLE & \CIRCLE & \CIRCLE & \CIRCLE & \CIRCLE & \CIRCLE & \CIRCLE & \CIRCLE & \CIRCLE   & \CIRCLE  & \Circle & \CIRCLE  & \CIRCLE  & \CIRCLE  & \CIRCLE  & \CIRCLE  & \CIRCLE  & \CIRCLE  & \Circle  & \CIRCLE & \CIRCLE\\

Free of charge (all) & \CIRCLE & \LEFTcircle & \LEFTcircle &  \CIRCLE & \Circle & \LEFTcircle & \LEFTcircle & \CIRCLE & \Circle & \CIRCLE & \Circle & \LEFTcircle & \Circle & \CIRCLE & \Circle & \Circle &\Circle & \Circle & \Circle & \CIRCLE & \Circle\\

\multicolumn{22}{c}{\cellcolor{grey!60}Legal bases} \\

Consent & \CIRCLE& \CIRCLE\zerowidthfootnote{Legal basis as such do not exist as a concept in the Australian Privacy Act.} & \CIRCLE& \CIRCLE& \CIRCLE& \CIRCLE& \CIRCLE& \CIRCLE& \CIRCLE& \CIRCLE& \CIRCLE& \CIRCLE& \CIRCLE& \CIRCLE& \CIRCLE& \CIRCLE& \CIRCLE& \CIRCLE& \CIRCLE& \CIRCLE& \CIRCLE\\  

\hspace{1em} $\hookrightarrow$ As the norm\footnote{Implying that the other legal basis should only be selected if there is a necessity.} & \CIRCLE & \Circle & \Circle & \Circle & \CIRCLE & \CIRCLE & \Circle &  \CIRCLE & \Circle & \Circle& \Circle &\Circle &  \CIRCLE &  \CIRCLE & \Circle  & \Circle &\Circle & \Circle & \CIRCLE & \Circle & \Circle\\

\hspace{1em} $\hookrightarrow$ Free & \CIRCLE & \CIRCLE& \CIRCLE& \CIRCLE& \CIRCLE& \CIRCLE& \CIRCLE& \CIRCLE& \CIRCLE& \CIRCLE& \CIRCLE& \CIRCLE& \CIRCLE& \CIRCLE& \CIRCLE & \CIRCLE & \CIRCLE & \CIRCLE & \CIRCLE & \CIRCLE & \CIRCLE \\

\hspace{1em} $\hookrightarrow$ Informed & \CIRCLE & \CIRCLE& \CIRCLE& \CIRCLE& \CIRCLE& \CIRCLE& \CIRCLE& \CIRCLE& \CIRCLE& \CIRCLE& \CIRCLE& \CIRCLE& \CIRCLE& \CIRCLE& \CIRCLE & \CIRCLE & \CIRCLE & \CIRCLE & \Circle & \CIRCLE & \CIRCLE \\

\hspace{1em} $\hookrightarrow$ Unambigous/explicit & \CIRCLE & \CIRCLE& \CIRCLE& \CIRCLE& \CIRCLE& \CIRCLE& \CIRCLE& \CIRCLE& \CIRCLE& \CIRCLE& \CIRCLE& \CIRCLE& \CIRCLE& \CIRCLE& \CIRCLE & \CIRCLE & \CIRCLE & \CIRCLE & \Circle & \CIRCLE & \CIRCLE \\

Legitimate interest & \Circle & \Circle & \Circle & \CIRCLE & \Circle & \CIRCLE  & \Circle & \Circle & \CIRCLE & \CIRCLE & \Circle & \CIRCLE & \Circle & \Circle & \CIRCLE  & \CIRCLE  & \CIRCLE  & \CIRCLE  & \CIRCLE & \CIRCLE &  \CIRCLE\\  

Contractual obligations & \CIRCLE  & \CIRCLE  & \Circle  & \CIRCLE & \Circle & \CIRCLE & \CIRCLE & \CIRCLE& \CIRCLE & \CIRCLE &\Circle & \CIRCLE& \CIRCLE& \CIRCLE& \Circle & \CIRCLE& \CIRCLE& \CIRCLE& \CIRCLE & \CIRCLE& \CIRCLE\\  

Legal obligation & \CIRCLE & \CIRCLE& \CIRCLE& \CIRCLE& \CIRCLE& \CIRCLE& \CIRCLE& \CIRCLE& \CIRCLE& \CIRCLE& \CIRCLE& \CIRCLE& \CIRCLE& \CIRCLE& \CIRCLE & \CIRCLE & \CIRCLE & \CIRCLE & \CIRCLE & \CIRCLE & \CIRCLE\\  

Vital interest & \Circle & \CIRCLE & \CIRCLE & \CIRCLE & \CIRCLE & \CIRCLE & \CIRCLE & \CIRCLE & \Circle& \CIRCLE & \CIRCLE & \CIRCLE  & \CIRCLE & \CIRCLE & \CIRCLE &\CIRCLE  & \CIRCLE & \CIRCLE & \CIRCLE & \CIRCLE  & \CIRCLE \\  

Public task & \CIRCLE & \CIRCLE & \Circle & \CIRCLE & \CIRCLE & \CIRCLE & \CIRCLE & \Circle & \CIRCLE & \CIRCLE & \CIRCLE & \CIRCLE & \CIRCLE & \Circle & \CIRCLE & \CIRCLE & \CIRCLE & \CIRCLE & \CIRCLE & \CIRCLE  & \CIRCLE\\  

Other legal basis for SP & \CIRCLE & \CIRCLE & \CIRCLE & \CIRCLE & \CIRCLE & \CIRCLE & \CIRCLE & \CIRCLE & \CIRCLE & \CIRCLE & \Circle & \CIRCLE & \CIRCLE & \CIRCLE & \CIRCLE & \Circle & \CIRCLE& \CIRCLE& \CIRCLE& \CIRCLE & \CIRCLE\\  

\multicolumn{22}{c}{\cellcolor{grey!60}Actors} \\

Data subject & \CIRCLE & \CIRCLE & \CIRCLE & \CIRCLE &\CIRCLE & \CIRCLE & \CIRCLE & \CIRCLE & \CIRCLE & \CIRCLE & \CIRCLE & \CIRCLE & \CIRCLE & \CIRCLE & \CIRCLE & \CIRCLE & \CIRCLE & \CIRCLE & \CIRCLE & \CIRCLE   & \CIRCLE\\  

Controller & \CIRCLE & \LEFTcircle\zerowidthfootnote{Both Canada's an Australia's federal data protection law do not distinguish between controller and processor, and even have very different conceptualization compared to the GDPR \cite{dataguidance_gdpr_pipeda,dataguidance_gdpr_australia}. As such, we coded is a partially similar.} & \CIRCLE & \CIRCLE  & \LEFTcircle & \CIRCLE & \CIRCLE &\CIRCLE &\CIRCLE &\CIRCLE &\CIRCLE &\CIRCLE &\CIRCLE &\CIRCLE &\CIRCLE &\CIRCLE &\CIRCLE &\CIRCLE &\CIRCLE &\CIRCLE & \CIRCLE\\  

Processor & \Circle & \Circle & \CIRCLE & \CIRCLE & \Circle & \CIRCLE & \CIRCLE & \CIRCLE & \CIRCLE & \CIRCLE & \CIRCLE & \CIRCLE & \CIRCLE & \CIRCLE\zerowidthfootnote{Different concpetualization, which can be translated as something as 'entrusted persons'.} & \CIRCLE & \CIRCLE & \CIRCLE & \CIRCLE & \CIRCLE & \CIRCLE & \CIRCLE \\  

Data Protection Authority
& \CIRCLE & \CIRCLE & \CIRCLE & \CIRCLE &\CIRCLE & \CIRCLE & \CIRCLE & \CIRCLE & \CIRCLE & \CIRCLE & \CIRCLE & \CIRCLE & \CIRCLE & \CIRCLE & \CIRCLE & \CIRCLE & \CIRCLE & \CIRCLE & \CIRCLE & \CIRCLE & \CIRCLE\\  

Data protection officer & \LEFTcircle & \LEFTcircle & \Circle  & \CIRCLE & \CIRCLE & \CIRCLE & \CIRCLE  & \CIRCLE  & \CIRCLE  & \CIRCLE  & \CIRCLE & \LEFTcircle  & \CIRCLE & \CIRCLE & \CIRCLE & \CIRCLE & \CIRCLE & \CIRCLE  & \CIRCLE  & \CIRCLE & \CIRCLE\\  

\multicolumn{22}{c}{\cellcolor{grey!60}Minors/Children} \\

In the DP law & \Circle & \Circle & \Circle & \CIRCLE & \Circle & \CIRCLE & \CIRCLE & \CIRCLE& \CIRCLE & \CIRCLE & \CIRCLE & \CIRCLE & \Circle & \Circle & \Circle &\Circle & \CIRCLE & \CIRCLE & \CIRCLE & \Circle & \CIRCLE\\  

\multicolumn{22}{c}{\cellcolor{grey!60}Transborder flows} \\

Allowed per conditions & \CIRCLE  & \CIRCLE  & \CIRCLE  & \CIRCLE  & \CIRCLE  & \CIRCLE  & \CIRCLE  & \CIRCLE  & \CIRCLE  & \CIRCLE  & \CIRCLE  & \CIRCLE  & \CIRCLE  & \CIRCLE  & \CIRCLE  & \CIRCLE  & \CIRCLE  & \CIRCLE  & \CIRCLE & \CIRCLE & \CIRCLE\\  

SCC & \CIRCLE & \CIRCLE  & \Circle & \CIRCLE & \Circle & \CIRCLE & \CIRCLE & \CIRCLE & \Circle & \CIRCLE  &\Circle & \Circle & \Circle & \Circle  & \CIRCLE &\Circle & \Circle & \Circle & \Circle &   \CIRCLE & \CIRCLE\\  

BCR & \CIRCLE & \CIRCLE & \Circle & \CIRCLE & \Circle & \CIRCLE & \Circle & \CIRCLE & \Circle & \Circle &\Circle & \Circle & \Circle & \CIRCLE & \CIRCLE & \CIRCLE & \CIRCLE & \Circle & \Circle & \CIRCLE & \CIRCLE\\  

Consent (common) & \CIRCLE & \CIRCLE & \CIRCLE & \CIRCLE & \Circle & \Circle & \Circle & \CIRCLE & \Circle & \Circle &\Circle &\Circle & \CIRCLE & \LEFTcircle\zerowidthfootnote{Data transfers outside Mexico do not require the consent of the data subject, but rather inform them that it is happening \cite{DataGuidanceMexico}.} & \CIRCLE & \CIRCLE & \CIRCLE & \CIRCLE & \Circle & \Circle & \Circle \\  

Consent (derogation)\footnote{Understood as it should only be used if no other reason is accepted, and not on the same level.} & \LEFTcircle\zerowidthfootnote{Only for medical data - Art.12(2)(b).} & \Circle & \Circle & \Circle & \Circle & \CIRCLE & \Circle & \Circle & \CIRCLE\zerowidthfootnote{We could not identify if it was common or not, so we err on the more stringent side.} & \CIRCLE &\Circle & \CIRCLE &\Circle& \Circle & \Circle &\Circle & \Circle & \Circle & \CIRCLE & \CIRCLE & \CIRCLE \\  

Adequacy list & \Circle & \Circle & \Circle & \CIRCLE & \Circle & \CIRCLE & \Circle & \CIRCLE & \CIRCLE & \CIRCLE &\Circle & \LEFTcircle\zerowidthfootnote{Per our interpretation of the law and based on \cite{DLAPiper2026IndonesiaDataProtection}} & \CIRCLE& \Circle & \Circle &\LEFTcircle\zerowidthfootnote{The adequacy list is based on the APEC's CBPR certification, which is not exactly the same conceptualization of our codebook \cite{singapore2021pdpr}.} & \Circle & \CIRCLE & \Circle & \CIRCLE & \CIRCLE\\  

\multicolumn{22}{c}{\cellcolor{grey!60}Security requirements for RDPR} \\

Security requirements & \CIRCLE  & \CIRCLE  & \CIRCLE  & \CIRCLE  & \CIRCLE  & \CIRCLE  & \CIRCLE  & \CIRCLE  & \CIRCLE  & \CIRCLE  & \CIRCLE  & \CIRCLE  & \CIRCLE  & \CIRCLE  & \CIRCLE  & \CIRCLE  & \CIRCLE  & \CIRCLE  & \CIRCLE & \CIRCLE & \CIRCLE\\  

Data breach definition & \Circle& \CIRCLE   & \Circle & \CIRCLE & \CIRCLE & \CIRCLE & \CIRCLE & \Circle & \CIRCLE & \CIRCLE & \CIRCLE & \CIRCLE & \CIRCLE & \Circle & \Circle & \CIRCLE & \CIRCLE & \Circle & \Circle & \Circle & \CIRCLE\\  

Notification & \Circle  & \CIRCLE  & \CIRCLE  & \CIRCLE  & \CIRCLE  & \CIRCLE  & \CIRCLE  & \CIRCLE  & \CIRCLE  & \CIRCLE  & \CIRCLE  & \CIRCLE  & \CIRCLE  & \CIRCLE  & \CIRCLE  & \CIRCLE  & \CIRCLE  & \CIRCLE  & \CIRCLE & \CIRCLE & \CIRCLE\\

\hspace{1em} $\hookrightarrow$ Authority & \Circle& \CIRCLE & \Circle & \CIRCLE & \CIRCLE & \CIRCLE & \CIRCLE & \Circle & \Circle & \CIRCLE & \CIRCLE & \CIRCLE & \CIRCLE & \Circle & \CIRCLE & \CIRCLE & \CIRCLE & \CIRCLE & \CIRCLE & \CIRCLE & \CIRCLE\\  

\hspace{1em} $\hookrightarrow$ Data subject & \Circle& \CIRCLE & \CIRCLE & \CIRCLE & \CIRCLE & \CIRCLE & \CIRCLE & \Circle & \CIRCLE & \CIRCLE & \CIRCLE & \CIRCLE & \CIRCLE & \CIRCLE & \CIRCLE & \CIRCLE & \CIRCLE & \CIRCLE & \Circle & \CIRCLE & \CIRCLE\\  

\hspace{1em} $\hookrightarrow$ Time frames & \Circle & \CIRCLE & \Circle & \CIRCLE & \CIRCLE & \CIRCLE & \Circle & \CIRCLE & \CIRCLE & \CIRCLE & \Circle & \CIRCLE & \CIRCLE & \CIRCLE & \CIRCLE & \CIRCLE & \CIRCLE & \CIRCLE & \CIRCLE & \CIRCLE & \CIRCLE\\  

Fine & \CIRCLE & \CIRCLE  &  \Circle  & \CIRCLE  & \CIRCLE  & \CIRCLE  & \CIRCLE  & \CIRCLE  & \CIRCLE  & \CIRCLE  & \CIRCLE  & \CIRCLE  & \CIRCLE  & \CIRCLE  & \CIRCLE  & \CIRCLE  & \CIRCLE  & \CIRCLE  & \CIRCLE & \CIRCLE & \CIRCLE\\  

Training &\Circle & \Circle & \Circle & \CIRCLE & \CIRCLE & \CIRCLE & \CIRCLE & \Circle & \CIRCLE & \CIRCLE & \CIRCLE & \CIRCLE & \CIRCLE & \CIRCLE & \Circle & \CIRCLE & \CIRCLE & \CIRCLE & \Circle & \CIRCLE & \CIRCLE\\  

\multicolumn{22}{c}{\cellcolor{grey!60}Documentation} \\

Privacy policy & \CIRCLE  & \CIRCLE  & \CIRCLE  & \CIRCLE  & \CIRCLE  & \CIRCLE  & \CIRCLE  & \CIRCLE  & \CIRCLE  & \CIRCLE  & \CIRCLE  & \CIRCLE  & \CIRCLE  & \CIRCLE  & \CIRCLE  & \CIRCLE  & \CIRCLE  & \CIRCLE  & \CIRCLE & \CIRCLE & \CIRCLE\\  

Privacy Notification& \CIRCLE  & \CIRCLE  & \CIRCLE  & \CIRCLE  & \CIRCLE  & \CIRCLE  & \CIRCLE  & \CIRCLE  & \CIRCLE  & \CIRCLE  & \CIRCLE  & \CIRCLE  & \CIRCLE  & \CIRCLE  & \CIRCLE  & \CIRCLE  & \CIRCLE  & \CIRCLE  & \CIRCLE & \CIRCLE & \CIRCLE\\  

RPA & \Circle & \Circle & \CIRCLE & \CIRCLE & \Circle & \CIRCLE & \CIRCLE & \CIRCLE & \CIRCLE & \CIRCLE &\Circle & \CIRCLE & \Circle& \Circle & \CIRCLE &\Circle & \CIRCLE & \CIRCLE & \Circle & \Circle & \CIRCLE\\  

DPIA & \LEFTcircle & \Circle & \Circle & \CIRCLE & \Circle & \CIRCLE & \CIRCLE & \Circle & \CIRCLE & \CIRCLE & \CIRCLE & \CIRCLE & \CIRCLE\zerowidthfootnote{Focused into security aspects.} & \Circle & \CIRCLE & \CIRCLE & \CIRCLE & \CIRCLE & \Circle & \LEFTcircle & \CIRCLE\\  

Registration authorities$^\dagger$ & \Circle & \Circle & \CIRCLE & \Circle & \Circle & \CIRCLE & \Circle & \CIRCLE & \CIRCLE & \Circle &\Circle & \Circle & \Circle & \Circle & \Circle &\Circle & \LEFTcircle & \CIRCLE & \CIRCLE & \CIRCLE & \CIRCLE\\  

\multicolumn{22}{c}{\cellcolor{grey!60}Privacy Notice Information} \\

\begin{tabular}{@{}l@{}}Identity and contact details of \\ \,\, controller\end{tabular} & \CIRCLE  & \CIRCLE  & \CIRCLE  & \CIRCLE  & \CIRCLE  & \CIRCLE  & \CIRCLE  & \CIRCLE  & \CIRCLE  & \CIRCLE  & \Circle  & \CIRCLE  & \CIRCLE  & \CIRCLE  & \CIRCLE  & \CIRCLE  & \CIRCLE  & \Circle  & \Circle & \CIRCLE & \CIRCLE\\  

DPO contact details & \Circle & \Circle & \Circle & \Circle & \Circle & \CIRCLE & \Circle & \Circle & \CIRCLE & \CIRCLE &\Circle & \Circle & \Circle & \Circle & \Circle &\Circle & \Circle & \Circle & \Circle & \Circle & \CIRCLE\\  

 Purpose and legal basis & \CIRCLE  & \CIRCLE  & \CIRCLE  & \CIRCLE  & \CIRCLE  & \CIRCLE  & \CIRCLE  & \CIRCLE  & \CIRCLE  & \CIRCLE  & \CIRCLE  & \CIRCLE  & \CIRCLE  & \CIRCLE  & \CIRCLE  & \CIRCLE  & \CIRCLE  & \CIRCLE  & \CIRCLE & \CIRCLE & \CIRCLE\\  

Legitimate interest explanation  & \Circle & \Circle & \Circle & \Circle & \Circle & \CIRCLE & \Circle & \Circle & \CIRCLE & \Circle &\Circle & \Circle & \Circle & \Circle & \Circle &\Circle & \Circle & \Circle & \Circle & \Circle & \CIRCLE\\

\begin{tabular}{@{}l@{}}Recipient or categories of \\ \,\, recipients of personal data\end{tabular} & \CIRCLE  & \CIRCLE  & \CIRCLE  & \CIRCLE  & \Circle   & \CIRCLE  & \Circle  & \Circle  & \CIRCLE  & \CIRCLE  & \Circle & \Circle  & \Circle  & \Circle  & \CIRCLE  & \Circle & \CIRCLE  & \CIRCLE  & \CIRCLE & \CIRCLE & \CIRCLE\\

International transfers information & \Circle  & \CIRCLE  & \CIRCLE  & \Circle  & \Circle  & \CIRCLE  & \Circle  & \Circle  & \CIRCLE  & \CIRCLE  & \Circle & \Circle  & \Circle  & \Circle  & \Circle  & \Circle & \CIRCLE  & \Circle  & \Circle & \Circle & \CIRCLE\\

Retention time & \Circle  & \Circle & \Circle  & \CIRCLE  & \Circle  & \CIRCLE  & \CIRCLE  & \Circle  & \CIRCLE  & \CIRCLE  & \Circle & \Circle  & \Circle  & \Circle  & \Circle  & \Circle & \Circle  & \CIRCLE  & \Circle & \Circle & \CIRCLE\\

\begin{tabular}{@{}l@{}}Explain legal or \\ \,\, contractual obligations\end{tabular} & \CIRCLE  & \CIRCLE & \CIRCLE  & \Circle & \Circle  & \Circle & \Circle  & \Circle  & \Circle & \CIRCLE  & \Circle & \Circle  & \Circle  & \Circle  & \Circle  & \Circle & \Circle  & \Circle  & \Circle & \Circle & \CIRCLE\\

Explanation of DSR & \CIRCLE  & \CIRCLE  & \CIRCLE  & \CIRCLE  & \Circle & \CIRCLE  & \CIRCLE  & \CIRCLE  & \CIRCLE  & \CIRCLE  & \CIRCLE  & \Circle  & \CIRCLE  & \CIRCLE  & \CIRCLE  & \Circle  & \CIRCLE  & \Circle & \CIRCLE & \CIRCLE & \CIRCLE\\

\hspace{1em} $\hookrightarrow$ General provision & \CIRCLE  & \Circle & \CIRCLE  & \CIRCLE & \Circle  & \Circle & \CIRCLE  & \CIRCLE  & \CIRCLE & \Circle  & \Circle & \Circle  & \CIRCLE  & \CIRCLE  & \CIRCLE & \Circle & \Circle  & \Circle  & \Circle & \CIRCLE & \Circle\\

\hspace{1em} $\hookrightarrow$ Detailed provision & \Circle  & \Circle & \Circle & \Circle & \Circle  & \CIRCLE & \Circle  & \Circle & \Circle & \CIRCLE  & \Circle & \Circle  & \Circle  & \Circle & \Circle & \Circle & \Circle  & \Circle  & \Circle & \Circle & \CIRCLE\\

\hspace{1em} $\hookrightarrow$ Specific rights & \Circle  & \CIRCLE & \Circle & \Circle & \Circle  & \Circle & \Circle  & \Circle & \Circle & \Circle  & \CIRCLE & \Circle  & \Circle  & \Circle & \Circle & \Circle & \CIRCLE  & \Circle  & \CIRCLE & \Circle & \Circle\\

\hline

\end{longtable}
\label{tab:regulation}
\end{landscape}

\section{Original spanish quotes} \label{app:original-quotes}

\textit{``Dato personal es, no recuerdo de memoria completamente, pero es cualquier dato que identifique o haga identificable a una persona. Prácticamente es como la esencia y no varía mucho de lo que vemos en otras latitudes del mundo'' (P 49).
}

\textit{``Está anclado en la idea del consentimiento y, lamentablemente, no hay esta incorporación de otras bases legales o están incorporadas de una forma que demanda un cierto ejercicio interpretativo'' (P 54).}

\textit{``Por ejemplo, el artículo 17 sobre derecho al olvido, eso es una patraña que no tiene pies ni cabeza en América Latina. Y no es que América Latina no contemple el derecho de cancelación'' (P 49).}

\textit{``No se puede hacer data compliance si no tienes data governance y lo mismo viceversa, no puedes hacer privacy si no tienes cybersec'' (P 68).}

\textit{``Y luego la reacción debiera hacer la evaluación de riesgo respecto a qué es lo que se filtró. Si la evaluación de riesgo determina que, por ejemplo, hay una afectación a privacidad, hay una afectación a datos personales sensibles, o a categorías especiales de datos, hay que tomar las medidas que la ley establece en este caso, por ejemplo, la notificación de la agencia'' (P 57).}

\textit{``Los datos de niños y niñas adolescentes, el proyecto de ley establece reglas especiales, lo asimila, pero no completamente a la categoría de datos personales sensibles'' (P 57)};

\textit{``No tenemos eso, no tenemos derecho al olvido, pero la Corte Constitucional, vía jurisprudencia, sí ha reconocido ese derecho, y dice, esto es una ramita, el derecho a supresión de datos'' (P 67)}

\textit{``Habiendo dicho esto, creo que las regulaciones tienen muchos puntos en común con el GDPR, pero también creo que existen muchas diferencias y existen cosas que no se pueden importar. Por ejemplo, el artículo 17 sobre derecho al olvido, eso es una patraña que no tiene pies ni cabeza en América Latina. Y no es que América Latina no contemple el derecho de cancelación. Existe el derecho de cancelación, pero es muy diferente al derecho de supresión y es muy diferente a que tú pongas a una empresa como juez y parte para determinar si algo tiene que mantenerse o no en línea, y sobre todo si no eres tú quien lo creo. Esto surge mucho por el tema de la indexación de información ya existente en Internet por parte de motores de búsqueda, que en su momento un caso contra Google, pero en América Latina nuestras realidades es que ¡Uf!Censura !'' (P 49)}.

\textit{``Interés legítimo no existe de ninguna forma'' (P 54)}.

\textit{``El régimen actual porque solo está el consentimiento, no hay más. Del resto solo es un régimen de excepciones, entonces es complicadísimo acá poder obtener el dato, o sea, si no tienes el consentimiento, el titular, chao'' (P 67)}.

\textit{``Creo que los puedes inferir a partir de la protección de datos sensibles con la enumeración de los datos que hace'' (P 55)}.

\textit{``Our regulations regarding countries that our country considers suitable, oh surprise, are practically the same as Europe’s. And, in fact, we added England at the same time as Europe; well, n... of course, Europe had added it earlier, and we did so later, so, well, we’re keeping up and following their footsteps'' (P 68)}.

\section{DPO stories} \label{annex:user-stories}
\newcounter{dpoid}
\setcounter{dpoid}{0}
\newcommand{\dpoid}{\stepcounter{dpoid}\arabic{dpoid}}

\begin{longtblr}[
  caption={DPO stories},
  label={table:user_stories},
]{
  colspec={lXlQ[l,2cm]},
  rowsep = {0pt},
  cells = {font=\scriptsize},
  cell{header} = {c=4}{c,gray!60},
  cell{group} = {c=4}{c,gray!20},
  hline{2} = {},
  rowhead=1,
}
ID & User Story & BAIT & SDLC \\
\SetChild{class=header} COMMON RDPRs && \\
\SetChild{class=group} Personal data && \\
\dpoid & As a DPO, I want to determine whether a data point can lead to the identification to a living natural person as personal data, so that I can apply the appropriate regulatory data protection requirements to it. & Information & Goals \\
\dpoid & As a DPO, I want to assess the risk of re-identification of anonymized data, so that I can determine if it remains outside the scope of regulatory data protection requirements. & Information & Testing \\
\dpoid & As a DPO, I want to identify the purpose behind the processing of each personal data point, so that I can limit the processing of that data for only that purpose. & Information & Requirements, Architecture, Design, Implementation \\
\dpoid & As a DPO, I want to keep a record of the processing purpose for each personal data point, so that I can demonstrate compliance with the principle of purpose limitation. & Business & Requirements \\
\dpoid & As an organization, I want to determine whether a data point belongs to a special category, so that I can identify the corresponding additional regulatory data protection requirements for its processing & Information & Requirements \\
\dpoid & As a DPO, I want to determine whether the purposes for processing special-category data can be achieved without using special-category data, so that I can minimize the use of special categories of data. & Business & Goals \\
\dpoid & As a DPO, I want to keep a record of the justification for processing special-categories of data, so that the organization can comply with the data minimization design principle. & Business & Design \\
\dpoid & As a DPO, I want to identify data protection requirements for processing special-category data specific to the jurisdiction the organization is subject to, so that I can ensure the organizational processes comply with them.  & Information & Requirements \\ 
\SetChild{class=group} Actors && \\
\dpoid & As an organization, I want to identify my role in each personal data processing activity, so that I can determine the specific legal obligation applicable to my role. & Business & Requirements \\
\dpoid & As an organization, I want to appoint a data protection officer, so that I can ensure oversight of data protection compliance. & Business & Operation \\
\dpoid & As a controller, I want to establish a contract with the processor defining each party's tasks and responsibilities, so that the data processing is conducted lawfully. & Business & Goals \\
\dpoid & As a controller, I want to include termination clauses in the controller–processor contract, so that data processing obligations are clearly defined upon contract termination. & Business & Requirements \\
\dpoid & As a controller, I want to specify in the controller-processor contract whether personal data must be deleted or returned upon termination, so that data processing obligations are clear at the point of contract termination. & Business & Operation \\
\dpoid & As a controller, I want to verify that personal data has been deleted or returned by the processor when a contract ends, so that the personal data cannot be reused.   & Business & Operation \\
\dpoid & As an organization, I want to determine the jurisdictions I am subject to, so that I can identify the relevant data protection authorities and comply with their requirements. & Business & Goals \\
\dpoid & As a controller, I want to hold exclusive responsibility for determining the purposes of each personal data processing activity, so that accountability for purpose definition is unambiguously assigned. & Business & Goals \\
\dpoid & As a controller, I want to review the security measures of the processor, so that I can ensure they provide sufficient guarantees to protect personal data. & Business & Operation \\
\dpoid & As a controller, I want to identify any third parties involved in the personal data processing, so that I can map where the personal data is being shared. & Application & Operation \\
\dpoid & As a processor, I want to process the personal data only based on the controller's documented instructions, so that I ensure compliance with contractual and regulatory obligations. & Business & Operation \\
\dpoid & As a processor, I want to assist the controller with data subject requests, so that the controller can fulfill its obligations regarding data subject rights under the applicable data protection regulations. & Business & Operation \\
\dpoid & As a data subject, I want to know which actors are involved in the processing of my personal data, so that I can understand who has access to my data. & Application & Operation \\
\dpoid & As an organization, I want to record when I am notified that a data subject has deceased, so that I can determine whether data protection requirements continues to apply to their personal data. & Information & Operation \\ 

\SetChild{class=group} Legal bases && \\
\dpoid & As an DPO, I want to gather consent in a freely given, informed, specific, and unambiguous manner, so that I can comply with regulatory requirements. & Information & Requirements \\
\dpoid & As an organization, I want to maintain verifiable evidence that consent was obtained in a clear, informed and transparent manner, so that I can demonstrate compliance with consent requirements. & Business & Requirements \\
\dpoid & As a data subject, I want to be informed when the purpose for asking my consent has changed, so that I can confirm whether my consent remains valid for the new purpose. & Business & Operation \\
\dpoid & As a data subject, I want to withdraw my consent for a personal data processing in a simple and accessible manner, so that I can stop the processing of my personal data by the organization.  & Application & Design, Operation \\
\dpoid & As an organization, I want to cease the processing of subject's personal data when consent is withdrawn, so that I can comply with legal obligations to stop processing upon withdrawal. & Information & Operation \\
\dpoid & As a DPO, I want to identify how local authorities interpret consent requirements, so that I can ensure the organization complies with the applicable guidance. & Business & Requirements \\
\dpoid & As a DPO, I want to identify when processing personal data is necessary for the performance of a contract to which the data subject is a party, so that I can establish the contractual legal basis for that processing activity. & Business & Requirements \\
\dpoid & As a DPO, I want to identify sector-specific legislations that may mandate the processing of specific personal data points, so that I can establish legal obligation as the lawful basis for the processing. & Business & Requirements \\
\dpoid & As a DPO, I want to identify the specific public interest that justifies the personal data processing, so that I can justify the legal basis.  & Business & Requirements \\
\dpoid & As a DPO, I want to identify when a data subject life may be at risk, so that I can process their personal data under the legal basis vital interest.   & Business & Requirements \\
\dpoid & As an organization, I want to delete personal data that does not have a legal ground for the processing, so that I can avoid unlawful processing. & Business & Operation \\

\SetChild{class=group} Data subjects rights && \\
\dpoid & As a DPO, I want to define processes to handle data subject rights request.  & Business & Goals \\
\dpoid & As an organization, I want to provide data subjects with a privacy notice explaining how their personal data is processed, so that they can make an informed decision about their personal data and rights. & Application & Requirements \\
\dpoid & As an organization, I want to have my privacy notice accessible to data subjects. & Application & Design \\
\dpoid & As a DPO, I want to write the privacy notice in plain language, so that data subjects can understand what they read.  & Application & Requirements \\
\dpoid & As a data subject, I want to be able to contact the data protection officer of an organization, so that I can exercise my data protection rights. & Information & Operation \\
\dpoid & As a data subject, I want to receive a free copy of my personal data that an organization holds about me, so that I can know all the information an organization has on me. & Information & Operation \\
\dpoid & As an organization, I want to gather all the personal data of a data subject within a defined time-frame including the data shared with third parties, so that I can provide a complete and timely copy of their data to the data subject.  & Information & Requirements, Architecture \\
\dpoid & As a data subject, I want to contact an organization when I believe my personal data is inaccurate, so that they can rectify it and keep accurate information. & Information & Operation \\
\dpoid & As a DPO, I want to identify inaccurate personal data across all the systems, so that I can rectify it as soon as it is noticed.  & Information & Operation \\
\dpoid & As an organization, I want to detect when processing of personal data has achieved its defined purposes, so that I can delete the data and comply with the purpose limitation principle.   & Information & Operation \\
\dpoid & As a data subject, I want to request the deletion of my data when I believe it no longer serves a purpose. & Information & Operation \\
\dpoid & As a data subject, I want to object the processing of my personal data by an organization. & Information & Operation \\
\dpoid & As a data subject, I want to restrict the processing of my personal data by an organization. & Information & Operation \\
\dpoid & As an organization, I want to temporarily restrict the processing of a data subject's personal data when a restriction/objection is presented, so that I can resolve the dispute.  & Information & Operation \\
\dpoid & As an organization, I want to keep a record of all data subjects rights request and their outcomes, so that I can demonstrate compliance with data protection authorities.  & Information & Operation \\ 

\SetChild{class=group} Minor's personal data && \\
\dpoid & As an organization, I want to identify what is the local legal definition of a minor, so that I can determine whether I am processing personal data from minors.  & Business & Requirements \\
\dpoid & As an organization, I want to identify what are the local regulatory requirements applicable to the processing of minor's personal data, so that I can satisfy them.  & Information & Requirements \\
\dpoid & As an organization, I want to classify the processing of minor's personal data as high risk, so that I can implement corresponding requirements applicable to the processing of high-risk personal data. & Information & Requirements \\
\dpoid & As an organization, I want to identify the requirements for guardian involvement when processing minor's personal data, so that I can ensure that the requirements are met. & Information & Requirements \\

\SetChild{class=group} Documentation && \\*
\dpoid & As an organization, I want to establish a privacy policy that defines our data protection governance framework. & Business & Goals \\
\dpoid & As an organization, I want to provide employees with an easily accessible privacy policy, so that I can keep them informed about the organization's privacy framework. & Application & Design \\
\dpoid & As a DPO, I want to have the privacy notice of the organization to include all the information required by the local data protection law. & Application & Requirements \\
\dpoid & As a data subject, I want to easily access the privacy notice written in plain language, so that I can understand its content. & Application & Requirements \\
\dpoid & As a DPO, I want to initiate a DPIA before starting a likely high-risk personal data processing activity, so that risks are identified and assessed prior to processing. & Information & Requirements \\
\dpoid & As a DPO, I want to document the outcomes of a DPIA, so that the organi`ation has an auditable record of the risk assessment. & Business & Goals \\
\dpoid & As a DPO, I want to implement mitigation measures for each risk identified by the DPIA. & Application & Architecture, Design, Implementation, Testing \\
\dpoid & As an organization, I want to keep a record of data breaches. & Information & Operation \\
\\ 
\\
\\
\\
\SetChild{class=header} UNCOMMON RDPRs&& \\
\SetChild{class=group} Time Frames && \\*
\dpoid & As an organization, I want to identify local time-frame for reporting a data breach, so that I can report them in time. & Business & Requirements \\
\dpoid & As an organization, I want to identify the local time-frames for providing each data subject right, so that I can fulfil them in time. & Information & Requirements \\
\dpoid & As an organzation, I want to identify all the time-frame requirements related to data protection. & Business & Requirements \\

\SetChild{class=group} Data Subject Rights && \\
\dpoid & As an organization, I want to identify how the right to cancellation/erasure is conceptualization in the jurisdiction I am subject to, so that I can understand the limitations/exceptions of this right. & Information & Requirements \\

\dpoid & As an organization, I want to determine if the right to portability is applicable in the jurisdiction I am subject to, so that I can establish the corresponding obligations for providing personal data in a machine-readable format. & Information & Requirements \\

\dpoid & As a DPO,  I want to identify any data subject rights recognised in the jurisdiction I am subject to beyond those in the baseline framework, so that I can define the requirements to satisfy them. & Information & Requirements \\

\dpoid & As a DPO, I want to identify conflicts between data subject rights requirements across the jurisdictions to which the organisation is subject. & Information & Requirements \\

\SetChild{class=group} Legal Bases && \\
\dpoid & As an organization, I want to identify the legal basis for processing personal data defined in the local data protection regulation, so that I can determine which are available for my processing activities. & Business & Requirements \\

\dpoid & As an organization, I want to identify legal bases for processing personal data established in sector-specific legislation applicable in the jurisdiction I am subject to, so that I can determine which are available for my processing activities. & Business & Requirements \\

\dpoid & As a DPO, I want to assess whether the organisation's legitimate interest overrides the rights and freedoms of the data subject, so that I can determine whether legitimate interest is a lawful basis for the processing activity. & Business & Requirements \\

\dpoid & As a DPO, I want to identify if non-consent legal bases are treated as exceptions in the jurisdiction I am subject to, so that I can determine when to ask for consent to data subjects. & Business & Requirements \\

\SetChild{class=group} Legal Bases && \\*
\dpoid & As an organization, I want to verify how vulnerability is conceptualized in the jurisdiction I am subject to, so that  I can asses if my information system interacts with such stakeholder. & Business & Requirements \\

\dpoid & As a DPO, I want to identify the data protection requirements when my information systems interacts with vulnerable actors. & Application & Requirements \\

\dpoid & As a DPO, I want to assess if there is a power asymmetry between actors as conceptualized in the jurisdiction I am subject to, so that I identify data protection requirements. & Business & Requirements \\

\end{longtblr}

\end{document}